\documentclass[a4paper]{article}
\usepackage{arxiv}
\usepackage[utf8]{inputenc}
\usepackage[T1]{fontenc}
\usepackage{hyperref}
\usepackage{url}
\usepackage{booktabs}
\usepackage{amsfonts}
\usepackage{nicefrac}
\usepackage{microtype}
\usepackage{lipsum}
\usepackage{graphicx}
\usepackage{doi}
\usepackage{enumerate}
\usepackage{nameref}
\usepackage{changepage}
\usepackage{comment}

\usepackage{caption}
\usepackage{subcaption}
\usepackage{natbib}

\usepackage{algorithm}
\usepackage[noend]{algpseudocode}
\usepackage{multirow}
\usepackage{adjustbox}
\usepackage{hhline}

\usepackage[right]{lineno}
\usepackage{setspace}
\usepackage{makecell}
\raggedright
\usepackage{lastpage,fancyhdr}
\usepackage{epstopdf}
\usepackage{color}
\definecolor{Gray}{gray}{.25}
\usepackage{sidecap}
\usepackage{wrapfig}
\usepackage[pscoord]{eso-pic}
\usepackage[fulladjust]{marginnote}

\newcommand{\footaffil}[2]{
  \begingroup
  \renewcommand*{\thefootnote}{\fnsymbol{footnote}}
  \footnotetext[#1]{#2}
  \endgroup
}

\title{Online Posting Effects: Unveiling the Non-linear Journeys of Users in Depression Communities on Reddit}

\author{
  \Large Virginia Morini\textsuperscript{ *†1},\footaffil{1},
  Salvatore Citraro\textsuperscript{ †1},\footaffil{2},
  Elena Sajno\textsuperscript{ *‡2},\footaffil{3},
  Maria Sansoni\textsuperscript{ §2},\footaffil{4} \\ \ \\ \ \\
  \Large \textbf{Giuseppe Riva}\textsuperscript{ ¶},\footaffil{5},
  \textbf{Massimo Stella}\textsuperscript{ ||3},\footaffil{6},
  \textbf{Giulio Rossetti}\textsuperscript{ †3}
}

\begin{document}
\justifying
\maketitle

\footaffil{1}{ Department of Computer Science, University of Pisa, Pisa, Italy}
\footaffil{2}{ Institute of Information Science and Technologies “A. Faedo” (ISTI), National Research Council (CNR), Italy}
\footaffil{3}{ Humane Technology Lab, Catholic University of Sacred Heart, Milano, Italy}
\footaffil{4}{ Department of Psychology, Catholic University of Sacred Heart, Milano, Italy}
\footaffil{5}{ Applied Technology for Neuro-Psychology Lab, IRCCS Istituto Auxologico Italiano, Italy}
\footaffil{6}{ CogNosco Lab, Department of Psychology and Cognitive Science, University of Trento, Italy, Email: massimo.stella-1@unitn.it}

\noindent \textbf{Author contributions}: V.M., S.C., M.St., G.Ro. designed research; V.M., S.C., G.Ro. performed research and analyzed data; M.S., M. St., E.S., G.R. provided psychological interpretations; All authors wrote the paper. \\ \ \\
\noindent \textsuperscript{1,2,3} V.M. and S.C.; E.S. and M.S.; M.St. and G.Ro. provided equal contributions.
\\ \ \\ \ \\

\begin{abstract}
Social media platforms have become pivotal as self-help forums, enabling individuals to share personal experiences and seek support. However, on topics as sensitive as depression, what are the consequences of online self-disclosure? 
 Here, we delve into the dynamics of mental health discourse on various Reddit boards focused on depression. To this aim, we introduce a data-informed framework reconstructing online dynamics from 303k users interacting over two years.
Through user-generated content, we identify 4 distinct psychological states resembling the stages of the Patient Health Engagement model. Our longitudinal analysis unveils online posting effects: a user can transition to another psychological state after online exposure to peers' emotional/semantic content.
As described by conditional Markov chains and different levels of social exposure, users' transitions reveal non-linear journeys with individuals navigating through both positive and negative phases in a spiral rather than a linear progression.  Interpreted in light of psychological literature, our findings provide evidence that the type and layout of online social interactions can deeply influence users’ journeys when posting about depression.
\end{abstract}

\section{Introduction}
\label{sec:intro}
Approximately one out of 26 people suffered from depressive symptoms in their lives \citep{gbd2019}.
Depression can deeply affect how people feel, think, act and communicate \citep{clark2000scientific}. As a condition of negative affect, depression can correspond to feelings of hopelessness, devaluation of life, anhedonia and self-deprecation, among others \citep{lovibond1995structure}. 
Traditionally, support for depression primarily involved direct, in-person interactions, where the physical presence of a therapist or support group could provide tangible, empathetic assistance \citep{clark2000scientific,naslund2016future}.
In the last few decades, the World Wide Web has shifted much of peer support to online social platforms (OSPs) \citep{escobar2018passive}. These digital spaces provide a venue where individuals can anonymously share their experiences and seek support, thus expanding access to assistance beyond conventional therapeutic environments \citep{deduro2024}.

The importance of seeking social support is well documented, as these interactions can play a critical role in the recovery process by building a support network, validating feelings, and sharing coping strategies \citep{garssen2021does,lazarus1984stress}.
 Indeed, online self-help groups offer the advantage of reaching potentially larger audiences and allowing for interaction without time and space constraints. This translates into immediate help and a constant availability that traditional settings cannot match \citep{naslund2016future,joseph2023cognitive}.
While this online shift has made seeking help for depression more accessible,
it has also introduced complexities relative to how support can be provided and received, thus raising an important question: \emph{Do online discussions benefit people debating depression?}

This research question can be explored at multiple levels, some more challenging than others. One way could be to quantify 
people's ``benefit'' in terms of understanding how online social interactions influence users' perceived depression symptoms over time. However, exploring this aspect at the individual level requires external validation coming from mental health professionals. The latter would have to follow
online users' patient journeys with depression, i.e. the fluctuations of mood and well-being relative to coping or dealing with depressive symptoms \citep{graffigna2013make,lovibond1995structure}. Whereas online data on OSPs is relatively easily available, e.g., Reddit counts thousands of posts in depression-related communities \citep{joseph2023cognitive}, external validation from experts is scarcely available, mainly because of anonymity, confidentiality, and lack of diagnosis \citep{ma2016anonymity}. This methodological challenge creates a significant gap in understanding how online discussions affect users' mental well-being, particularly given the massive scale of online mental health communities \citep{pantic2014online,tang2021emotional}. To address this gap, researchers have disregarded individual-based questionnaires and relied on artificial intelligence (AI) methods to map digital traces (e.g., users' posts or likes) \citep{stella2022cognitive} in mental well-being dimensions of distress. Most of these studies \citep{escobar2018passive,fatima2019prediction,fatima2021dasentimental} usually focus on predicting distress levels within individual posts, one by one, crucially neglecting social or time-wise interactions among users.

We argue that users in OSPs cannot be reduced to sequences of likes or posts but are rather complex systems entwining cognitive content, emotional perspectives, and mental well-being dimensions. To tackle all these directions at once, we tap into theoretical frameworks spanning cognitive data science \citep{stella2022cognitive}, clinical psychology \citep{clark2000scientific,graffigna2013make}, psycholinguistics \citep{tausczik2010psychological}, natural language processing \citep{joseph2023cognitive} and social network analysis \citep{rossetti2018community}.
Focusing on the Reddit platform, here we investigate the above complexities by harnessing whether online debates on depression can correspond to alterations, either positive or negative, in users' expressed well-being over time. \newline
To this aim, in the following we briefly discuss key aspects of OSPs and their role in mental health self-disclosure and psychological content engagement.

\subsection*{Self-disclosure on OSPs like Reddit}
In psychotherapy, disclosing personal experiences for seeking mental health support is a well-established practice \citep{clark2000scientific,holahan2005stress}. Whereas therapy traditionally occurs between the therapist and the client, nowadays, OSPs have quickly become self-help communities where users share personal information with peers facing similar challenges \citep{pantic2014online,huffaker2010dimensions,tang2021emotional}. These exchanges often involve conversations or discussions among more than two users at a time, all protected by the anonymity OSPs offer \citep{naslund2016future}. Recent literature discusses how anonymity, due to the online disinhibition effect \citep{ma2016anonymity}, can make it easier to disclose personal information with strangers rather than with personal acquaintances.

Among OSPs, Reddit is a social platform where users post content to forums called \textit{subreddits}, each dedicated to specific topics or interests.
Also Reddit allows almost complete anonymity -- unless users choose otherwise -- enabling discussions about sensitive issues and disclosures of personal experiences \citep{ma2016anonymity}. These exchanges are often moderated in ways that every user could be comfortable enough to share their own thoughts in a safe environment without toxic social interactions \citep{almerekhi2020}.
Furthermore, users can either provide or seek support, thus impersonating more than one role in the same community (e.g., caregiver or help-seeker) \citep{balsamo2023}. Beyond communicative roles and their relative intentions, e.g., seeking support or providing trust, users on Reddit can engage in different threads over time and potentially undergo some inner changes. In other words, users can change or evolve in their communicative intentions over time after exposure to online content or personal happenings in their offline lives. Despite the latter events occurring outside the online world, users might still report events in their lives within Reddit boards, ultimately using subreddits as personal diaries \citep{abramski2024}. 

\subsection*{The psychology of content engagement}
Using Reddit communities as personal diaries crucially enriches the online environment with a wide variety of real-world and personal events. In turn, this richness increases the chances that online content might appeal, trigger or resonate with other users, potentially going through similar or disparate events \citep{almerekhi2020}. 
Ultimately, people might react in different ways after being exposed to such rich digital content via online social interactions.
Accordingly, past works have found that even reading online content can be a powerful experience for someone's affect and cognition \citep{kramer2014experimental, goldenberg2020digital, joseph2023cognitive}. This phenomenon, known as \textit{emotional contagion}, refers to the act of any two individuals exchanging knowledge and converging towards the same emotional state \citep{hatfield1993emotional}. Emotional contagion was found to occur also in the absence of face-to-face interactions: Even reading online posts \citep{kramer2014experimental} can alter the emotional states of people and the effect lasts for hours \citep{ferrara2015measuring}.
As a consequence, engaging with different types of content can elicit different reactions in users participating in the same discussion thread. 

\textit{How do users express their own reactions?} On Reddit, users can express their thoughts mostly via posts, endorsements, replies, and- re-sharings, leaving digital traces that can thus be considered as proxies into users' expressed psychological state at the moment of writing \citep{huffaker2010dimensions,joseph2023cognitive,fatima2019prediction}. It is important to underline that digital traces cannot fully reconstruct the psychology of an online user. Instead, these traces represent proxies that can partially reconstruct what users are talking about and how they emotionally perceive something \citep{stella2022cognitive}. These two elements map semantic- and emotional content of users' psychology and could potentially be used to identify clusters of users sharing similar psychological features. This passage is supported by rich psychological literature establishing a link between people's language and their psychological features. 

In general, the Deep Lexical Hypothesis \citep{uher2013personality} posits that psychological constructs and traits can percolate through one's psychology up to alter people's language \citep{cutler2022deep}. Focusing on depression, several studies have identified that individuals with higher depressive symptoms exhibit the following features in their language: (i) absolutist thinking \citep{al2018absolute} (e.g., a black-and-white vision of the world), (ii) self-focused language \citep{rude2004language,al2018absolute} (e.g., mentioning the self disproportionately), (iii) negative emotions \citep{tausczik2010psychological,stephan2013enactive} (e.g., overabundance of anger and fear) and (iv) a low sense of dominance and agency \citep{stephan2013enactive} (e.g. an inability for someone to feel in control of their life events). Hence, these features can help in discerning different levels or categories of users debating and affected by depression to different extents.

These features of posts and/or comments, merged with their underlying social interactions, provide a complex yet rich snapshot of online social debate about depression. In other words, we argue that next-generation computational social science studies should go beyond modeling social networks as ``skeletons'' of who-answers-to-whom and rather embrace the psychological, cognitive, and time-evolving nature of online human interactions, especially within online supportive communities. 

\subsection*{Research challenges and study aims}
Following this reasoning, this study focus on understanding how online social interactions taking place on Reddit  affect individuals discussing depression. While prior research has explored depression detection on social media \citep{escobar2018passive,kamarudin2021study,fatima2019prediction}, the psychological impact of these interactions remains poorly understood. Here, we propose a novel framework that combines psycholinguistic analysis with social network dynamics to investigate how users' psychological states evolve through online interactions.

Unlike previous studies, we examine how users' language and social connections jointly reflect their psychological journey. While acknowledging that someone's language may not overlap entirely with their mental state \citep{gentzkow2019text}, our approach combines psycholinguistic features extracted from users' content, social interaction patterns captured through network analysis, and temporal dynamics modeled via conditional Markov chains.\newline
This methodological framework addresses our main research question:
\begin{itemize}
\item[\textbf{RQ1.}] \textit{How do online interactions within support communities impact individuals' psychological states?}
\end{itemize}
To fully investigate this general question, we further explore two specific aspects:

\smallskip
\noindent\textbf{RQ1.1.} Do individuals' psychological states form distinct patterns that could align with established psychological models?

\smallskip
\noindent\textbf{RQ1.2.} Do users progress linearly through these states or follow more complex patterns?

\smallskip
From our two-year analysis of 303K Reddit users in depression-related communities, we identified four data-driven psychological states. Interestingly, these states show notable parallels with established psychological frameworks, particularly the Patient Health Engagement (PHE) model \citep{graffigna2013make}. This model describes mental health management through phases of emotional overwhelm (\textit{blackout}), initial awareness (\textit{arousal}), emerging engagement (\textit{adesion}), and structured management (\textit{eudaimonic project}) -- a progression that offers valuable context for interpreting our findings, as we will discuss. 
Subsequently, we quantitatively reconstruct users' ``journeys'' through these states, introducing the concept of ``posting effects'' -- the likelihood of transitioning between psychological states after specific types of social exposure. 

Through this novel methodological lens, we find that online interactions can both improve and potentially worsen users' expressed mental well-being, challenging linear assumptions about recovery trajectories.

\section*{Method}
Our study leverages a longitudinal Reddit\footnote{https://reddit.com/} user interactions dataset. We chose Reddit as our data source for several key reasons: i) its community-focused structure through topic-specific subreddits is widely used for mental health support-seeking, enabling targeted analysis of depression-related discussions \citep{proferes2021studying}, ii) the platform's anonymity facilitates more open self-disclosure about mental health experiences, and iii) its threaded discussion format allows for systematic tracking of user interactions and conversation dynamics.
Given our focus on depression and following previous research \citep{fatima2019prediction,sampath2022data,escobar2018passive,kamarudin2021study}, we selected six popular subreddits related to depression in general, treatments, and support seeking (see Table \ref{tab:dataset_desc} for details about subscribers and activity metrics). We collected two years of data (May 1, 2018, to May 1, 2020) from these communities using the Pushshift API \citep{baumgartner2020pushshift}, resulting in 378,483 posts and 1,475,044 comments from 303,016 unique users\footnote{The anonymized data are made available in a dedicated \href{https://github.com/virgiiim/SocialEffects_Reddit_CaseStudy}{GitHub Repository}.}. To ensure data quality, we implemented several cleaning steps: filtering for English-language content, removing duplicated posts/comments, empty content, and deleted accounts, and excluding content from moderators and known Reddit bots\footnote{\url{https://botrank.pastimes.eu/}}. Since our focus is on user interactions, we retained only users who engaged with at least one other user. For each post and comment, we preserved essential metadata including pseudonymized author identifiers, content text, timestamps, and thread structure information. Additional details about the data collection steps are available in \textit{SI}, Section 1.

\begin{table*}[b]
\centering
\begin{tabular}{|c|c|c|c|c|}
\hline
\textbf{Subreddit} & \textbf{\# subscribers} & \textbf{\# posts} & \textbf{\# comments} & \textbf{\# users} \\ 
\hline
{\fontfamily{qcr}\selectfont r/depression} & 928,705 & \multirow{6}{*}{378,483} & \multirow{6}{*}{1,475,044} & \multirow{6}{*}{303,016} \\
\cline{1-2}
{\fontfamily{qcr}\selectfont r/depressionregimens} & 43,832 & & & \\
\cline{1-2}
{\fontfamily{qcr}\selectfont r/depression\_help} & 81,971 & & & \\
\cline{1-2}
{\fontfamily{qcr}\selectfont r/EOOD} & 89,686 & & & \\
\cline{1-2}
{\fontfamily{qcr}\selectfont r/GFD} & 11,807 & & & \\
\cline{1-2}
{\fontfamily{qcr}\selectfont r/sad} & 133,154 & & & \\
\hline
\end{tabular}
\caption{\textbf{Dataset Description}: For each considered subreddit: number of subscribers, the total number of posts and comments extracted, and the total number of users.}
\label{tab:dataset_desc}
\end{table*}

In the following, we describe our analytical pipeline, focusing on its main components, namely: (i) the definition of the User Generated Contents (UGC) features used to characterize the psychological dimensions of the selected population; (ii) the identification of users' psychological states; (iii) the measuring and validation of engagement and social exposure effects on online users' journeys.
\\ \ \\
\noindent{\bf Stage 1: Mapping UGCs Psycholinguistic dimensions.}
We extracted a set of features from users' posts/comments along 5 psychological and linguistic dimensions: 
\begin{itemize}
    \item[i.] \emph{Plutchik’s Primary Emotions} provides information about words' expressive level among eight basic emotions, according to the well-established Plutchik’s psychoevolutionary theory of primary emotions \citep{plutchik1980general}; in detail, we use the emotional ratings of the NRC Lexicon\footnote{\url{https://github.com/metalcorebear/NRCLex}} \citep{mohammad2013crowdsourcing};
    \item[ii.] \emph{PAD Emotional Dimensions} measures the level of valence/pleasure, arousal, and dominance associated with users' contents, which indicates, respectively, the level of pleasantness, stimulation, and control experienced by users based on the Mehrabian and Rusell PAD model \citep{mehrabian1974approach}; in detail, we use the NRC VAD Lexicon \citep{vad-acl2018};
    \item[iii.] \emph{Sentiment} unveils whether the underlying emotional tone of users’ texts appears to be positive/negative; in detail, we leverage the VADER Lexicon\footnote{\url{https://github.com/cjhutto/vaderSentiment}} \citep{hutto2014vader};
    \item[iv.] \emph{Taboo Rate} captures the frequency of taboo or offensive words within a given text \citep{reilly2020building};
    \item[v.] \emph{Subjectivity} refers to the level of opinions or personal feelings expressed in users' texts, as opposed to objective facts; in detail, we rely on the TextBlob library\footnote{\url{https://textblob.readthedocs.io/en/dev/}}.
\end{itemize}

We assessed the significance and non-redundancy of extracted features via statistical analysis (see \emph{SI}, Section 2).
\newline
To generate monthly scores for each user, we averaged the values of these indicators based on the content they shared. Additional details on feature extraction and preprocessing can be found in \textit{SI}, Section 2.
\\ \ \\
\noindent{\bf Step 2: Identifying Users' Psychological States and Discussed Topics.}
We used the extracted indicators to identify groups of similar users through unsupervised clustering.
Each user is thus represented as a vector of feature values proxying the user's psychological state during a given month.
We leverage the K-Means  \citep{macqueen1967classification} to identify four psychological clusters based on the retrieved psycholinguistic features. 
The parameter value choice, $k=4$, is driven by the elbow method, a well-known method used to estimate the best number of clusters present in a dataset.
Technical details of K-Means can be found in \textit{SI}, Section 2.

Regarding clusters' content analysis, we investigate all texts produced by users belonging to a cluster via topic modeling, i.e., an unsupervised learning method to automatically cluster groups of words that best characterize a set of documents. In detail, we rely on BERTopic \citep{devlin2018bert} (via Python implementation\footnote{\url{https://github.com/MaartenGr/BERTopic}}) to extract ten meaningful topics from each psycholinguistic cluster (see Figure 1(a-d) of \textit{SI}). 
Technical details of K-Means, and BERTopic algorithms can be found in \textit{SI}, Section 2.
\\ \ \\
\noindent{\bf Stage 3: Measuring Social Exposure.}
To measure the effects of social exposure on online users' journeys, we filtered our dataset. We kept only those users (along with their interactions temporally aggregated at the monthly level) having participated in depression-related discussions for at least two consecutive months.
After such a filter, the initial population was reduced by $50\%$ ($\simeq 150.000$ active users).
Leveraging such sub-population we defined different levels of social exposure and relate them to the likelihood of cluster change.
Fixed a month $t$, a generic user $u$ and a conditioning cluster $C$, we measure social exposure as a result of different levels of interactions:
\begin{itemize}
    \item \textit{Single Interaction}: during $t$, $u$ interacted with at least a user $v$ of cluster $C$ - where being in contact means that $u$ and $v$ directly reply each other in at least a discussion thread;
    \item \textit{Single Homogeneous Context}: during $t$, $u$ participated at least a \emph{discussion context} $\Gamma$ whose majority of users belong to cluster $C$ - where $\Gamma$ identifies a nested sub-thread (e.g., all messages appearing at the same level in the discussion tree of a given thread). Therefore, $\Gamma$ describes a set of users that, even non-replying directly to each other, participate in the same discussion.
    \item \textit{Majority of Interactions}: during $t$, $u$ interacted prevalently with users belonging to $C$;
    \item \textit{Majority of Homogeneous Contexts}: during $t$, $u$ participated prevalently in contexts where most users belong to cluster $C$.
\end{itemize}
For every exposure level, we compute the transition probability for a user to move from cluster $C_i$ at a generic time $t$ to cluster $C_j$ at time $t+1$.

Levels of social exposure are modeled on the network of interactions between users. To this aim, we leverage notions from graph theory, either pairwise graphs for the \textit{Single Interaction} and the \textit{Majority of Interactions} models \citep{newman2018networks}, or hypergraphs for the \textit{Single Homogeneous Context} and the \textit{Majority of Homogeneous Contexts} ones \citep{battiston2020networks}.
Direct replies between users were represented as pairwise edges between two graph nodes, whereas discussion contexts -- i.e., the sub-threads -- were represented using hyperedges, namely sets of nodes, of a hypergraph. 
On the one hand, two interacting users, i.e., replying to the same post, are linked through pairwise edges.
On the other hand, hypergraphs expand on the traditional graph structure by allowing for higher-order interactions between more than two nodes \citep{battiston2020networks}.
Users that rely on the same level of discussion jointly form a hyperedge.
Detailed references to the formal definitions and modeling choices used to capture pairwise and higher-order user interactions are reported in the SI, Section 3.

We use mathematical modeling to explore whether the above social interactions affect how individuals move across clusters over time. Specifically, we measure Constrained Transition Patterns (CTPs) as the conditional probability of observing individuals' transition from one cluster to another when users are exposed, to different degrees, to the content originating from a target cluster. Consequently, we obtain a matrix of Markov transition probabilities for every level of social exposure and constraining clusters. Note that with \textit{constraining cluster}, we intend the cluster upon which we are conditioning. 

We used two null models to statistically test the significance of those transitions. 
In doing so, we account for the following confounders:
\begin{itemize}
    \item[i.] the effect of the overall distribution of cluster labels - by destroying original users' cluster memberships;
    \item[ii.] the effect of time - by destroying the temporal dependency of connectivity between users.
\end{itemize}
Regarding the former confounder, clusters' labels of all users were randomly shuffled several times, complying with the original distribution.
Regarding the latter, monthly graph/hypergraph snapshots were randomly shuffled by keeping the original topologies.
Then, for both models, the expected probabilities to shift between clusters were compared to the probabilities computed on the real data.
Only statistically significant transitions ($p<0.01$) were kept in the analysis.
Refer to the SI, Section 3, for a more formal description of the null models and the results obtained by testing each null model independently.

 \section*{Results}
\label{sec:pre}
	
Our data-informed approach highlights users' transition patterns across online interactions: Users can embark on several possible journeys, transitioning across healing or turbulent paths. 
In the following, we report the main results provided by our analysis, focusing on i) the description of the identified users' clusters (i.e., psychological states characterizing groups of users), ii) the effects of social exposure as described by the statistically validated transition probabilities computed from observed online users' journeys. 

\subsection{Users' clusters as psychological states}
\begin{figure*}
    \centering
    \includegraphics[scale=0.43]{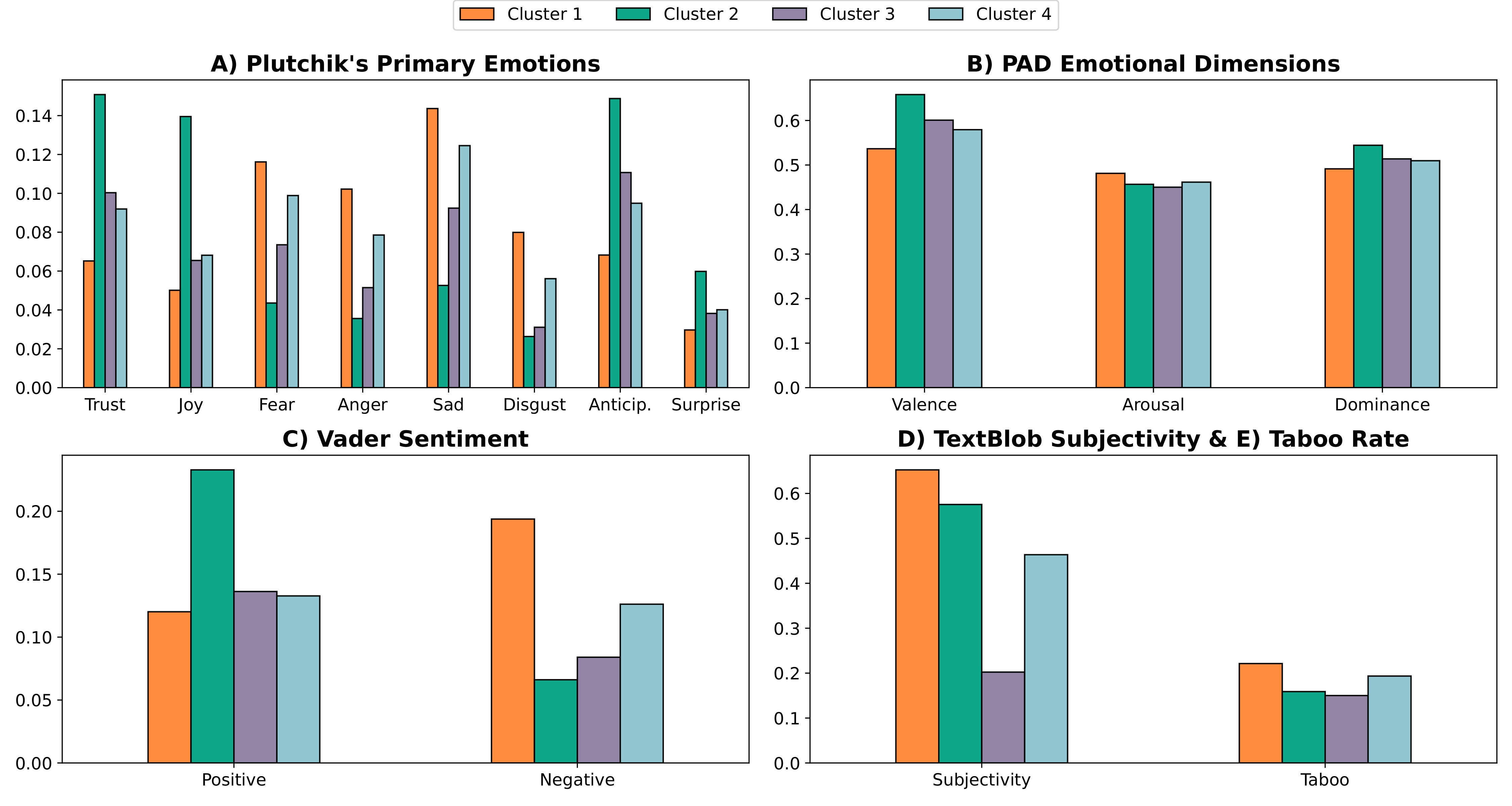}
    \caption{\textbf{Psycholinguistic cluster profiles in Reddit Depression discourse.} \textbf{(A-E)} Bar charts of cluster centroids values: \textit{Plutchik’s Primary Emotions},  \textit{PAD Emotional Dimensions}, \textit{VADER Sentiment}, \textit{Textblob Subjectivity}, \textit{Taboo Rate}.}
    \label{fig:cluster_features}
\end{figure*}

We characterize every cluster in terms of i) average features, i.e., K-Means cluster centroids as shown in Figure \ref{fig:cluster_features}, and ii) cluster content via topic detection using BERTopic, see \textit{SI} Figure 1(a-d). 
\\ \ \\
\noindent\textbf{Cluster 1: High Distress State.}
As shown in Figure \ref{fig:cluster_features}(A), Cluster 1 (C1) displays a highly negative emotional profile, with prominent levels of sadness, fear and disgust and an absence of joy or trust. C1 also has higher arousal but lower dominance and valence compared to other clusters (Figure \ref{fig:cluster_features}(B)), indicating alarm and anxiety \citep{mohammad2013crowdsourcing,russell1980circumplex} with lower self-agency \citep{vad-acl2018}. C1 posts also have the most negative VADER sentiment, a higher subjectivity (Figure \ref{fig:cluster_features}(C,D)) and many taboo words, indicating the presence of offensive and egocentric language (Figure \ref{fig:cluster_features}(E)). 

Topics in C1 (Figure 1(a) of \textit{SI}) include \textit{Education/Work}, \textit{Relationship}, \textit{Need Support}, \textit{Sleep Habits}, \textit{Appearance}, \textit{Meds/Doctor}, \textit{Family}, \textit{Suicide}, and \textit{Self-Harm}. 
The presence of topics around self-harming behaviors and suicidal thoughts among online users discussing depression mirrors offline expectations from the relevant clinical literature \citep{clark2000scientific}. Despite the negative profile, users seek support from others, enhancing emotional well-being \citep{keles2020systematic}. The negative and positive topics (Figure 1(a) of \textit{SI}), and overall negative profile (Figure \ref{fig:cluster_features}(A)) suggest ambivalence in C1 users' communicative intentions.
\\ \ \\
\noindent\textbf{Cluster 2: Positive Resilient State.}
In contrast to the highly negative emotional profile and distressed expressions of Custer 1, Cluster 2 (C2) demonstrates markedly different characteristics, suggesting a more positive psychological state.
Indeed, as shown in Figure \ref{fig:cluster_features}(A), C2 has higher levels of positive emotions like joy and trust compared to C1. Negative emotions and sentiment are less prevalent in C2 (Figure \ref{fig:cluster_features}(C)). C2 exhibits lower arousal, higher valence, and dominance (Figure \ref{fig:cluster_features}(N)), indicating positive activation and a sense of agency and control \citep{posner2005circumplex,vad-acl2018,stephan2013enactive}. These elements correspond to lower depression levels \citep{holahan2005stress,knapen2015exercise}. C2 has also lower taboo rates, suggesting non-aggressive communication, and high subjectivity, indicating a focus on personal perspectives \citep{gentzkow2019text} (Figure \ref{fig:cluster_features}(E,D)).

Topic analysis shows C2 users mention therapy, hobbies, sports, and medications (Figure 1(b) of \textit{SI}). Specifically,  Topics include \textit{Education/Work}, \textit{Family}, \textit{Relationship}, \textit{Friendship}, \textit{Need Support}, \textit{Therapy}, \textit{Happy moments}, \textit{Music}, and \textit{Pets}. \textit{Suicide} and \textit{Self-Harm} are absent. Interestingly, C2's topics and emotions relate to COPE model strategies like \textit{Instrumental/Emotional Social Support}, \textit{Positive Reinterpretation}, and \textit{Mental Disengagement} \citep{litman2006cope,litman2009frequency,lazarus1984stress}. These strategies reflect proactive emotion management and support-seeking from family, friends, therapy and hobbies. 
\\ \ \\
\noindent\textbf{Cluster 3: Balanced/Neutral State.}
While Cluster 2 exhibits predominantly positive emotions and proactive coping strategies, Cluster 3 (C3) presents a more balanced profile with both positive and negative elements. 
In fact, Cluster 3  combines sadness, trust and anticipation with average levels of arousal, valence and dominance (Figure \ref{fig:cluster_features}(A,B)). Users show a weakly positive VADER sentiment, a lower rate of taboo words and lower subjectivity compared to other clusters, thus indicating minimal personal bias and aggressiveness (Figure \ref{fig:cluster_features}(C-E)).

Despite these positive/neutral signals, topic modeling reveals that C3 focuses on technical depression-related terms, discussing topics like \textit{Education/Work}, \textit{Meds/Doctor}, \textit{Self-Harm} and  \textit{Suicide} (Figure 1(c) of \textit{SI})). 

These patterns indicate that C3 users employ both positive coping strategies, like \textit{Seeking Social Support}, and negative ones, like \textit{Self-Harm Coping} \citep{litman2006cope}. Despite a positive and proactive demeanor, support-seeking, and emphasis on therapy, the mention of \textit{Self-Harm} suggests internal conflicts.
\\ \ \\
\noindent\textbf{Cluster 4: Fluctuating State.}
Although Cluster 3 maintains a relatively balanced emotional profile, Cluster 4 (C4) shows more emotional variability, characterized by fluctuating patterns of both positive and negative expressions that distinguish it from the more stable nature of C3. Accordingly, as shown in Figure \ref{fig:cluster_features}(A), Cluster 4 (C4) users display more sadness, fear and anger and less anticipation than users in C3. C4 users also display moderate levels of arousal, valence and dominance (Figure \ref{fig:cluster_features}(B)). C4 users tend to display also more negative VADER sentiment, use more taboo words and display higher subjectivity compared to C3 (Figure \ref{fig:cluster_features}(C-E)).

Topic modelling reveals that users in C4 discuss \textit{Relationships}, \textit{Music}, \textit{Education/Work}, \textit{Friendship}, \textit{Pets}, and \textit{Weight} (Figure 1(d) of \textit{SI})). These are all activities and entities supporting mental well-being through social support \citep{krause2012pet,raglio2015effects}. However, professional and medical support is also mentioned in this cluster. Hence, C4 users might be trying to cope with depression by considering heightened arousal activities and coping strategies like mental disengagement. 

\begin{figure*}
    \centering
    \includegraphics[scale=0.125]{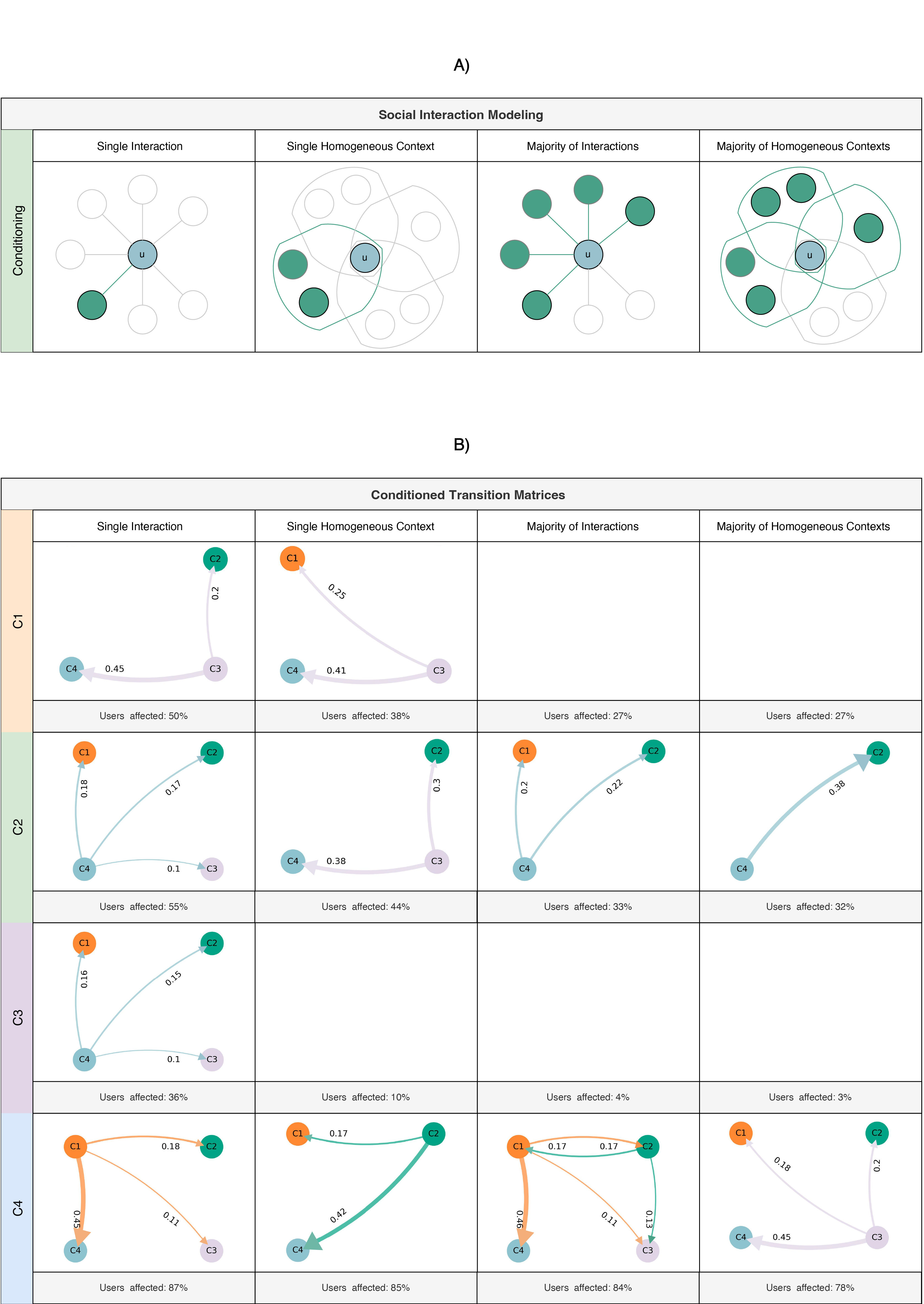}
  \caption{\textbf{Social influence and cluster transitions in Reddit Depression interactions.} \textbf{A)} Schematic examples characterizing the four modeled levels of social exposure. Nodes identify users (colored by their profile cluster), and edges/sets identify one-to-one interactions/discussion contexts. \textbf{B)} Visual representation of Markov transition matrices among user clusters conditioned on group-contact typology and filtered based on temporal and volume statistical significance. Rows identify the conditioning profile cluster, and columns the assumed interaction model. In each graph: nodes identify profile clusters; directed edges identify statistically significant transition given the conditioning social interaction with its observed probability; \emph{users affected} identify the percentage of the online population affected by the identified patterns.}
  \label{fig:dep_constr}
\end{figure*}
\subsection{Social interactions and online user journeys}
We model 4 levels of exposure to a constraining cluster as exemplified in Figure \ref{fig:dep_constr}(A). 
Single and majority of interactions/homogeneous contexts use pairwise links/hyperlinks to quantify the exposure to the target cluster. 
\textit{Single} means ``weak": there is at least one interaction with the target cluster in the users' social circles. \textit{Majority} means ``strong": the majority of interactions in users' social circles is with the target cluster. 
Figure \ref{fig:dep_constr}(B) reports the Markov process of the identified statistically significant CTPs. 

From the first row of Figure \ref{fig:dep_constr}(B), we can observe that conditioning over users in the \textit{High Distress State} \textit{C1} generates statistically significant transitions only for users in the \textit{Balanced/Neutral State} \textit{C3}.
Interestingly, in the single interaction model, users in the \textit{Balanced/Neutral State} \textit{C3} can transition to users in the \textit{Positive Resilient State} \textit{C2} only when interacting with users in the \textit{High Distress State} \textit{C1} (with a probability of $.20$).
However, in the single homogeneous context model, i.e., when a thread exposes users in the \textit{Balanced/Neutral State} \textit{C3} to a majority of peers in the \textit{High Distress State} \textit{C1}, they can transition into the \textit{High Distress State} \textit{C1} ($.25$) or into the \textit{Fluctuating State} \textit{C4} ($.41$).

Figure \ref{fig:dep_constr}(B) second-row underlines that exposure to users in the \textit{Positive Resilient State} \textit{C2} can lead to statistically significant transitions for users in both the \textit{Balanced/Neutral State} \textit{C3} and \textit{Fluctuating State} \textit{C4}.
Users in the \textit{Balanced/Neutral State} \textit{C3} show significant transitions only in the single homogeneous context model, where they can transition to the \textit{Positive Resilient State} \textit{C2} ($.30$) and the \textit{Fluctuating State} \textit{C4} ($.38$).
Users in the \textit{Fluctuating State} \textit{C4}, instead, are strongly affected by interactions with users in the \textit{Positive Resilient State} \textit{C2} across all exposure levels except for the single homogeneous context model. When users are exposed to content from users in the \textit{Positive Resilient State} \textit{C2} across all social threads/contexts, we observe a significant transition only for users in the \textit{Fluctuating State} \textit{C4} ($.38$). The same transition from users in the \textit{Fluctuating State} \textit{C4} to users in the \textit{Positive Resilient State} \textit{C2} is observed also in the single interaction model ($.17$) and in the majority of interactions model ($.22$).
Moreover, users in the \textit{Fluctuating State} \textit{C4} display a slight tendency toward moving to the \textit{Balanced/Neutral State} \textit{C3} ($.10$ in single interactions) and transitioning to the \textit{High Distress State} \textit{C1} ($.18$ in single interactions, $.20$ in the majority of interactions). These interactions indicate that being exposed to positive content can have a dualistic effect on users in the \textit{Fluctuating State} \textit{C4}.

Moving to the third row in Figure \ref{fig:dep_constr}(B), we observe that users in the \textit{Fluctuating State} \textit{C4} engaging with content from users in the \textit{Balanced/Neutral State} \textit{C3} can transition into all other states with analogous probabilities, $.16$ to the \textit{High Distress State} \textit{C1}, $.15$ to the \textit{Positive Resilient State} \textit{C2}, $.10$ to the \textit{Balanced/Neutral State} \textit{C3}.

Finally, Figure \ref{fig:dep_constr}(B) fourth-row describes the impact of interactions with users in the \textit{Fluctuating State} \textit{C4}. Since users in the \textit{Fluctuating State} \textit{C4} represent the most common profile in our longitudinal analysis, it is expected for them to influence most other users. This is reflected by a higher number of statistically significant transitions and the percentage of users affected reaching up to 87\%.
For instance, we observe that users in the \textit{Fluctuating State} \textit{C4} tend to maintain their state even after interacting with other users in the same state. 
Here, users in the \textit{Positive Resilient State} \textit{C2} can transition to other states only in single homogeneous contexts and in the majority of interactions scenarios. In both cases, users in the \textit{Positive Resilient State} \textit{C2} react to content from users in the \textit{Fluctuating State} \textit{C4} by transitioning to the \textit{High Distress State} \textit{C1} ($.17$ in single homogeneous contexts, and $.17$ in majority of interactions). 
Once in the \textit{High Distress State} \textit{C1}, users can transition back to the \textit{Positive Resilient State} \textit{C2} after being exposed to content from users in the \textit{Fluctuating State} \textit{C4}, even with single interactions ($.18$). This and other transitions (see Figure \ref{fig:dep_constr}(B)) contribute to a spiral indicating a back-and-forth transition between positive and negative psychological states.

\subsection*{Interpretation of users' transitions}
At the level of individual interactions, we find that exposure to content from users in the \textit{High Distress State} \textit{C1} influences only users in the \textit{Balanced/Neutral State} \textit{C3}.  
This is different from our expectations based on the well-known psychological mechanism of online emotional contagion: Kramer and colleagues \citep{kramer2014experimental} found that even reading an online post on Facebook can affect the emotional state of readers for a few hours. Instead, our transition findings indicate that the span of negative emotional contagion via single interactions on Reddit is not strong enough to move users discussing depression toward negative spirals. On the contrary, exposure to negative posts can even cause positive shifts, moving users from the \textit{Balanced/Neutral State} \textit{C3} to the \textit{Positive Resilient State} \textit{C2} 20\% of the time. This finding agrees with past studies \citep{niehoff2020know} reporting how people may value the sensations evoked by negative content, extracting social value from it while reducing their own uncertainty. 

Furthermore, single interactions with users in the \textit{Fluctuating State} \textit{C4} can move users from the \textit{High Distress State} \textit{C1} to the \textit{Fluctuating State} \textit{C4} almost $45\%$ of the time. This effect is almost 3 times stronger than most others observed within a single interaction and thus deserves more attention. We hypothesize such a transition reflects an emotional pattern of the patient journey \citep{rodriguez2023incorporation,graffigna2013make}, where individuals usually deal with pathological conditions, such as major depression, at first mostly through maladaptive/dysfunctional behavior but then move slowly toward acceptance while going through conflict and contrast. 

Going beyond pairwise social interactions, exposure to single homogenous contexts of users in the \textit{High Distress State} \textit{C1} does not affect users in the \textit{Positive Resilient State} \textit{C2}, indicating there might be a lack of direct flow between these states. This transition rate, compatible with random expectations, might provide quantitative confirmation of the patient journey \citep{rodriguez2023incorporation,graffigna2013make}, where moving from negative to positive affect is not instantaneous but rather requires journeying through intermediate emotional states. 

We also find that positive online interventions are globally ineffective in improving users' affect. Being surrounded by users in either the \textit{High Distress State} \textit{C1} or the \textit{Balanced/Neutral State} \textit{C3} provides no deviations from random expectations. Being surrounded by users in the \textit{Positive Resilient State} \textit{C2} is similarly ineffective since users in the \textit{Fluctuating State} \textit{C4} can transition either to the \textit{High Distress State} \textit{C1} or to the \textit{Positive Resilient State} \textit{C2} with roughly equal probabilities. Being surrounded by users in the \textit{Fluctuating State} \textit{C4} can affect users in the \textit{Positive Resilient State} \textit{C2} but to a considerably lesser extent when compared to homogenous/coordinated interactions. 
This difference further indicates that positive states are fragile to fluctuating interactions, which can even cause emotional volatility \citep{gee2020suicidal} i.e., direct transitions between users in the \textit{High Distress State} \textit{C1} and \textit{Positive Resilient State} \textit{C2}.
Exposure to content from users in the \textit{Fluctuating State} \textit{C4} can also massively affect users in the \textit{High Distress State} \textit{C1}, who can transition to less negative but still variable states (46\%).

\section*{Discussion}
In exploring how online interactions within support communities can impact users' psychological states, our study addresses two core questions: (1) firstly, whether these states exhibit distinct, identifiable patterns aligned with established psychological models; (2) secondly, whether users' progression through these states is linear or follows more complex trajectories.

Our study provides evidence that social interactions discussing depression online -- specifically on Reddit -- correlate with users' psychological states, as expressed in posts and narratives. To understand the psychological significance of these patterns, we return to the Patient Health Engagement (PHE) model \citep{graffigna2013make}, which offers a structured framework for analyzing how individuals progress through different stages of mental health awareness. Specifically, the user clusters detected in our analysis share notable parallels with the four stages described in the PHE model. Users in cluster C1, expressing a high distress state, demonstrate characteristics similar to the \textit{Blackout} phase of the PHE model. This phase typically manifests through emotional upheaval, a sense of turmoil, and a mix of venting and help-seeking behavior, reflecting the emotional fragility and ambivalence common in early phases of mental health awareness. Users in cluster C2, expressing a positive resilient state, represent individuals with a positive emotional disposition who engage in well-being-promoting activities and actively seek support. This pattern aligns with the \textit{Eudaimonic Project} phase, suggesting resilience and a constructive approach to mental health. We characterize this positive and resilient cluster C2 as the eudaimonically hopeful'' group, reflecting their optimism and proactive coping strategies. Users in cluster C3, expressing a balanced/neutral state, demonstrate a mental health phase marked by cognitive acceptance and adherence to medical support, though internal struggles persist. This corresponds to the \textit{Adhesion} phase of the PHE model, leading us to characterize C3 users as the medically adherent yet conflicting'' group. Users in cluster C4, expressing a fluctuating state, show heightened arousal and negative emotions, often with conflicting views, corresponding to the \textit{Arousal} stage in the PHE model.

In addressing our second research question, contrary to traditional offline engagement models \citep{graffigna2013make}, our findings indicate that online users navigate complex trajectories. Rather than following a direct path from turmoil to acceptance, users' psychological states fluctuate based on exposure to various types of online interactions, including individual replies and longer threads containing conflicting, adherent, or turbulent content exchanges. \newline
Firstly, we found an unexpected pattern regarding positive interactions. Single exchanges with hopeful/positive users from C2 correlate with state changes among users in conflicted states (C4). This pattern appears connected to C4's emotionally polarized nature, where users frequently express suicidal thoughts \citep{gee2020suicidal}. These expressions may indicate higher levels of suicide ideation \citep{teixeira2021revealing} among C4 users. Recent studies have shown how increased levels of suicide ideation correspond with greater emotional volatility \citep{gee2020suicidal}, reflecting a tendency to transition between extreme states of negative and positive affect. This volatility may explain why C4 users struggle to benefit from positive content and often revert to more negative psychological states. \newline
Secondly, we unveiled distinct patterns from individual exchanges when moving to group interactions. Indeed, interactions within homogeneous groups (e.g., discussion threads) showed stronger associations than single interactions, particularly regarding transitions from cluster C3 to C1. This pattern suggests that negative emotional contagion on Reddit requires sustained exposure rather than brief interactions, contrasting with findings from Facebook \citep{kramer2014experimental}. However, exposure to positive contexts showed mixed effectiveness, correlating with neutral transitions between conflict-ridden clusters (C3 and C4) approximately 38\% of the time. This finding underscores the complexity of the patient journey and highlights conflict's central role, even amid positive feedback, as previously hypothesized \citep{rodriguez2023incorporation,graffigna2013make,joseph2023cognitive}.
The psychological theory of the patient journey suggests that acceptance and hopefulness represent the final stage for individuals managing depression \citep{rodriguez2023incorporation,graffigna2013make}. However, our findings indicate that C2 does not function as an attractor state -- users can and do transition out of it. Exposure to conflicting homogeneous clusters from C4 correlates with transitions to C4 itself among C2 users about 42\% of the time. These interactions also associate with shifts from C2 to the negative state of C1 (17\% of cases). These patterns suggest that exposure to emotionally polarized and conflicting posts correlates with negative outcomes for users in positive psychological states, a concerning phenomenon warranting additional future research.

Notably, users in the most positive cluster show selective vulnerability -- they appear susceptible to conflicting posts while remaining relatively resilient to purely negative content. This difference may stem from two mechanisms. First, hopefulness and positive affect may enhance resilience to negative emotions, as suggested by the psychological theory developed by Tugade and colleagues \citep{tugade2004resilient}. Users might successfully process negative content but struggle with conflicting emotions, which require additional cognitive resources \citep{litman2006cope}. Alternatively, this pattern might reflect coping strategies of avoidance \citep{holahan2005stress}, where users maintain hopefulness by avoiding overtly negative content but remain vulnerable to content that appears less threatening yet contains emotional conflict. This phenomenon affects a substantial portion of the community, influencing over 85\% of the users on the depression board, warranting attention for future research on posting effects.

These findings demonstrate how integrating psychological content and social interactions can produce relevant psychological insights about the nature of conflict and online social engagement in depression-related subreddits.
\\ \ \\
\noindent\textbf{Practical Implications.} The findings of this work have significant implications for improving mental health support in online environments, informing both platform design, moderation, and clinical practice. 
First, our observations about non-linear psychological trajectories suggest potential insights for online support systems and mental health professionals. While established psychological frameworks -- such as the PHE model -- provide structured perspectives on mental health progression \citep{graffigna2013make}, our findings indicate that users' journeys through online support communities may involve varied patterns of engagement. This perspective could contribute to the development of support systems that consider the dynamic nature of online interactions. 

Moreover, the observed differences between individual and group interaction effects suggest potential adjustments to community structure and moderation approaches. Our results indicate that sustained exposure within discussion threads shows stronger associations with state changes than individual, one-to-one interactions. Platform designers might consider implementing tools to monitor the emotional content of discussion threads leveraging established theory on emotional contagion and user engagement on social media platforms \citep{kramer2014experimental,goldenberg2020digital,gross2015emotion}.

Additionally, the identified vulnerability of users in positive states to conflicting content points to the potential value of implementing protective features while maintaining user autonomy. Drawing from selective exposure theory \citep{Knobloch2014choice}, platforms could develop tools that help users manage their exposure to different types of content. Such features might include content filtering options and graduated engagement mechanisms that allow users to control their participation in more challenging discussions.

Finally, our findings suggest opportunities for enhancing moderator and peer supporter training. Community moderators and supporters should be trained to recognize different psychological states based on linguistic markers, understand the non-linear nature of recovery journeys, and implement evidence-based digital support strategies. This could improve the effectiveness of online support while minimizing potential negative impacts of well-intentioned but potentially destabilizing interactions.
\\ \ \\
\noindent\textbf{Limitations and Future Directions.}
This study acknowledges several  limitations that pave the way for future research. First, the lack of clinical diagnoses and offline contextual data for online users (such as real-world support networks, life events, or therapeutic interventions) presents challenges in accurately quantifying how users' content reflects distinct psychological constructs like depression, burnout, anxiety, and stress \citep{lovibond1995structure,koutsimani2019relationship,fatima2021dasentimental}. While obtaining such comprehensive data for large-scale studies involving hundreds of thousands of users would be impractical while maintaining privacy, our methodology aligns with recent advances in computational psychology that demonstrate how digital footprints can provide valuable insights when clinical data is unavailable \citep{fatima2019prediction,ferrara2015measuring,tausczik2010psychological}. Future research could strengthen this foundation by integrating additional psychometrically validated approaches.

Secondly, our focus on Reddit raises questions of generalizability. Yet the platform's key features - threaded discussions, community moderation, and anonymous participation - are common across mental health forums (e.g., TalkLife\footnote{https://talklife.com/}, 7 Cups\footnote{https://www.7cups.com/forum/}, Mental Health Forum\footnote{https://www.mentalhealthforum.net}), suggesting potential broader applicability that future studies should verify.

Additionally, two methodological considerations warrant attention. First, our approach of aggregating users over time periods potentially conflates individual emotional trajectories into single averages. Although this choice prioritized identifying global patterns in an unexplored online board, future studies should explore finer, time-based analyses of individual trajectories. Second, while mapping our clusters to PHE model stages \citep{graffigna2013make} is supported by quantitative methodology, it may oversimplify the complexity of psychological transitions, as these states exist on a continuum rather than as discrete categories.

Finally, our framework's language-agnostic approach to topic modeling, while offering versatility, may miss important context-specific nuances in depression-related discourse. The way certain concepts and ideas are perceived within mental health communities may differ significantly from their use in general language \citep{teixeira2021revealing}. The development of context-aware analysis frameworks could better account for depression-specific language use and its nuances in future research.

\section*{Declaration of Conflicting Interests}

The Authors declare that there is no conflict of interest.

\section*{Acknowledgments}
This work is supported by: the EU NextGenerationEU programme under the funding schemes PNRR-PE-AI FAIR (Future Artificial Intelligence Research); the EU – Horizon 2020 Program under the scheme ``INFRAIA-01-2018-2019 – Integrating Activities for Advanced Communities” (G.A. n.871042) ``SoBigData++: European Integrated Infrastructure for Social Mining and Big Data Analytics” (http://www.sobigdata.eu); PNRR-``SoBigData.it - Strengthening the Italian RI for Social Mining and Big Data Analytics'' - Prot. IR0000013. The authors thank Daniele Fadda and Eleonora Cappuccio for the visual analytics support.

{\Huge
\noindent \textbf{Supporting Information}}
\newline

\noindent This supplementary text outlines the methodological details of the work.
Section \ref{sec:collect} describes the data collection steps.
Section \ref{sec:user_profiling_meth} focuses on the features extracted from users' texts and on the grouping and analysis of these latter ones in precise cluster profiles.
Section \ref{sec:interacts} focuses on the networked representations used to quantify the social exposure between users and on the methodological details about the transition patterns between clusters.

\section{Data Collection}
\label{sec:collect}
Our data collection process, performed through the pushshift.io Reddit API\footnote{\url{https://github.com/pushshift/api}} \cite{baumgartner2020pushshift}, served two main purposes: i) gathering users' textual content (posts and comments) for psycholinguistic feature extraction and topic analysis (see Section \ref{sec:user_profiling_meth}), and ii) tracking user interactions in discussion threads to construct conversation networks (see Section \ref{sec:interacts}).
For each discussion thread, we collected both posts and associated comments. Posts represent the initial content that starts a discussion, while comments are replies either to the original post or to other comments, creating a threaded conversation structure. \newline
For each post, we extracted several key fields:
\begin{itemize}
   \item \textit{id}: unique post identifier
   \item \textit{author}: username
   \item \textit{selftext}: main content
   \item \textit{title}: post title
   \item \textit{created\_utc}: UNIX timestamp of creation
\end{itemize}
To capture the interactive nature of discussions, we collected the following fields for comments:
\begin{itemize}
   \item \textit{id}: unique comment identifier
   \item \textit{author}: username
   \item \textit{body}: comment text
   \item \textit{link\_id}: identifier of the parent post
   \item \textit{parent\_id}: identifier of the parent comment
   \item \textit{created\_utc}: UNIX timestamp of creation
\end{itemize}
The \textit{link\_id} and \textit{parent\_id} fields were particularly important as they allowed us to reconstruct the complete structure of each discussion thread, showing how users interacted with each other through their comments. To protect user privacy, we implemented pseudonymization during data collection: both \textit{id} and \textit{author} fields were processed through an irreversible hash function, ensuring that users cannot be identified while preserving our ability to track their interactions across the dataset.
\newline
The resulting dataset structure enables us to analyze both the content of users' communications and the patterns of their interactions. By combining these two aspects, we can examine how users engage with each other in depression-related discussions and how these interactions might influence their expressed psychological states over time.



\section{Clustering Profiles: Methods}
\label{sec:user_profiling_meth}

In the following, we first discuss the methodology we used to infer users' longitudinal (i.e., monthly) psycholinguistic profiles from written texts.
Then, we describe how we clustered and characterized four precise profiles.

\subsection{Extracting users' psycholinguistic profiles}
\label{sez:feature_extr}
From the texts (i.e., posts and comments) of each user, we extract a set of features along five psychological and linguistic dimensions: 
\begin{itemize}
    \item[-] \textbf{Plutchik’s Primary Emotions}: to identify what emotions are expressed by users,
    we refer to the well-established psychoevolutionary theory of emotions \cite{plutchik1980general} developed in 1980 by the psychologist Robert Plutchik. In detail,  Plutchik identified eight primary emotions - \textit{joy}, \textit{trust}, \textit{fear}, \textit{surprise}, \textit{sadness}, \textit{anticipation}, \textit{anger}, \textit{disgust} - and postulated that all other emotions are mixed or derivative states that occur as combinations, mixtures, or compounds of such eight basic emotion. To operationalize this theory in our study, we rely on the NRC (National Research Council) lexicon \cite{mohammad2013crowdsourcing}, a widely-used lexical resource that contains over 14,000 English words and their associated emotional ratings according to Plutchik’s theory of eight primary emotions. 
   In detail, for each user-generated text, we attribute emotions to individual words via the NRCLex python library\footnote{\url{https://github.com/metalcorebear/NRCLex}} that further expands the NRC affect lexicon from 14,000 to 27,000 words by including WordNet synonyms \cite{miller1995wordnet}. 
    In detail, emotion scores range from 0 to 1, where 0 means that words in a text do not elicit a certain affective association, and 1 represents that a text strongly associates with that affective category.
    
    \item[-] \textbf{PAD Emotional Dimensions}: to understand users' attitudes and behavioral intentions while writing on the platform, we refer to the psychological Pleasure, Arousal, Dominance (PAD) model introduced by the psychologist Albert Mehrabian and James A. Russell in 1974 \cite{mehrabian1974approach}. Specifically, Mehrabian and Russell  presented the idea of using three emotional dimensions - \textit{pleasure}/\textit{valence}, \textit{arousal}, \textit{dominance} - to describe perceptions of the environments where the individual is in. Accordingly, pleasure/valence deals with whether the individual perceives the environment as enjoyable or not. Arousal reflects the extent to which the environment
    stimulates the individual. Dominance captures whether
    the individual feels in control or not in the environment. In the current study, we operationalize the PAD model by leveraging the NRC Valence, Arousal, and Dominance (VAD) Lexicon \cite{vad-acl2018}. This lexicon consists of a comprehensive list of over 20,000 English words, along with their corresponding valence/pleasure, arousal, and dominance scores.
    The lexicon assigns to each word considered a score that ranges from 0 (lowest V/A/D) to 1 (highest V/A/D) for each of the three dimensions (V/A/D).  
    
    \item[-] \textbf{VADER Sentiment}: to unveil whether the underlying emotional tone of users' texts appears to be positive/negative, we leverage VADER (Valence Aware Dictionary and sEntiment Reasoner) \cite{hutto2014vader}, a lexicon and rule-based sentiment analysis tool tailored to deal with sentiments expressed in social media (e.g., short and dirty piece of texts). Specifically, we use the vaderSentiment Python library\footnote{\url{https://github.com/cjhutto/vaderSentiment}} to assign to each piece of text a \textit{positive} and \textit{negative} sentiment score in a range from 0 to 1, where 1 means the highest intensity of that sentiment and 0 the lowest. 

    \item[-] \textbf{Taboo Rate}: to capture whether users' discussions contain a potentially offensive language, we refer to the \textit{taboo rate}, i.e., the frequency of taboo or offensive words within a given text. Specifically, it quantifies the level of explicit or sensitive language present, which may include profanity, racial slurs, or other socially unacceptable terms. To obtain such an indicator, we refer to the Tabooness dataset\footnote{\url{https://www.reilly-coglab.com/data}} taken from \cite{reilly2020building} which consists of 1,205 high-frequency English words annotated with their taboo index. High values refer to offensive words, while low values to neutral ones.

    \item[-] \textbf{TextBlob Subjectivity}: to identify the degree of personal opinion, emotion, or bias present in users' texts, we refer to the concept of \textit{subjectivity}. Indeed, the subjectivity of a text is a measure of the level of personal feelings expressed in a text, as opposed to objective facts. To compute users' subjectivity we rely on the TextBlob Python library\footnote{\url{https://textblob.readthedocs.io/en/dev/}} that assigns a subjectivity score to a piece of text on a scale of 0 to 1, with 0 being completely objective and 1 being completely subjective. 
        
\end{itemize}
Users' longitudinal psycholinguistic profiles  (i.e., users' profiles for each month) are thus obtained by averaging values of the above indicators for each content they shared in a given month. 
Notice that before computing these features, we apply a standard text preprocessing pipeline to each post and comment, which involves converting text to lowercase, removing non-printable characters, XSLT tags, newline and tab characters, URLs, punctuation, and extra spaces. 
Additionally, we apply further text preprocessing to individual features as per the requirements of libraries or datasets. 
For a detailed description of these indicators, please refer to Table \ref{tab:feature}.
\begin{table}
\scriptsize
\centering
\caption{For each psychological and/or linguistic dimension, the textual features considered, feature range, and feature description.}
\label{tab:feature}
\begin{tabular}{|l|c|c|l|} 
\hline
\textbf{Dimension}  & \textbf{Features}     & \textbf{Range}                                                                                                    & \textbf{Description}                                                                            \\ 
\hhline{|====|}
                    & \textit{Anger}        & \textcolor[rgb]{0.122,0.125,0.141}{[}0\textcolor[rgb]{0.122,0.125,0.141}{,}1\textcolor[rgb]{0.122,0.125,0.141}{]} & \makecell[lt]{The degree of strong displeasure, hostility, and desire \\ for aggression.}  \\ 
\cline{2-4}
                    & \textit{Anticipation} & \textcolor[rgb]{0.122,0.125,0.141}{[}0\textcolor[rgb]{0.122,0.125,0.141}{,}1\textcolor[rgb]{0.122,0.125,0.141}{]} & \makecell[lt]{The degree of excitement, eagerness, and readiness for \\ future events or experiences.}                          \\ 
\cline{2-4}
                    & \textit{Disgust}      & \textcolor[rgb]{0.122,0.125,0.141}{[}0\textcolor[rgb]{0.122,0.125,0.141}{,}1\textcolor[rgb]{0.122,0.125,0.141}{]} & \makecell[lt]{The degree of revulsion, aversion, and strong distaste \\ towards something unpleasant or offensive.}             \\ 
\cline{2-4}
\makecell{\textbf{Plutchik’s}\textit{ }\\ \textbf{Primary} \\ \textbf{Emotions}}    & \textit{Fear}         & \textcolor[rgb]{0.122,0.125,0.141}{[}0\textcolor[rgb]{0.122,0.125,0.141}{,}1\textcolor[rgb]{0.122,0.125,0.141}{]} & \makecell[lt]{The degree of apprehension, anxiety, and protective \\ response to perceived danger or threat.  }                 \\ 
\cline{2-4}
                    & \textit{Joy}          & \textcolor[rgb]{0.122,0.125,0.141}{[}0\textcolor[rgb]{0.122,0.125,0.141}{,}1\textcolor[rgb]{0.122,0.125,0.141}{]} & \makecell[lt]{The degree of intense happiness, delight, and sense of \\ well-being.}                                           \\ 
\cline{2-4}
                    & \textit{Sadness}      & \textcolor[rgb]{0.122,0.125,0.141}{[}0\textcolor[rgb]{0.122,0.125,0.141}{,}1\textcolor[rgb]{0.122,0.125,0.141}{]} & \makecell[lt]{The degree of sorrow, grief,  and unhappiness often \\ accompanied by withdrawal and low mood.}                   \\ 
\cline{2-4}
                    & \textit{Surprise}     & \textcolor[rgb]{0.122,0.125,0.141}{[}0\textcolor[rgb]{0.122,0.125,0.141}{,}1\textcolor[rgb]{0.122,0.125,0.141}{]} & \makecell[lt]{The degree of astonishment, amazement, or disbelief \\ in response to the unexpected.}                            \\ 
\cline{2-4}
                    & \textit{Trust}        & \textcolor[rgb]{0.122,0.125,0.141}{[}0\textcolor[rgb]{0.122,0.125,0.141}{,}1\textcolor[rgb]{0.122,0.125,0.141}{]} & \makecell[lt]{The degree of confidence, reliance, and positive \\ expectations in others' honesty and intentions.}              \\ 
\hline
                    & \textit{Valence}      & \textcolor[rgb]{0.122,0.125,0.141}{[}0\textcolor[rgb]{0.122,0.125,0.141}{,}1\textcolor[rgb]{0.122,0.125,0.141}{]} & \makecell[lt]{The degree of pleasantness or unpleasantness \\ experienced.}~                                      \\ 
\cline{2-4}
\makecell{\textbf{PAD} \\ \textbf{Emotional} \\ \textbf{Dimensions}} & \textit{Arousal}      & \textcolor[rgb]{0.122,0.125,0.141}{[}0\textcolor[rgb]{0.122,0.125,0.141}{,}1\textcolor[rgb]{0.122,0.125,0.141}{]} & \makecell[lt]{The degree of excitement, stimulation,  or energy \\ experienced.}~                                  \\ 
\cline{2-4}
\makecell{\\}
                    & \textit{Dominance}    & \textcolor[rgb]{0.122,0.125,0.141}{[}0\textcolor[rgb]{0.122,0.125,0.141}{,}1\textcolor[rgb]{0.122,0.125,0.141}{]} & \makecell[lt]{The degree of influence or control perceived.}~                          \\ 
\hline
\makecell{\textbf{VADER} \\ \textbf{Sentiment}}  & \textit{Positive}     & \textcolor[rgb]{0.122,0.125,0.141}{[}0\textcolor[rgb]{0.122,0.125,0.141}{,}1\textcolor[rgb]{0.122,0.125,0.141}{]} & \makecell[lt]{The level of favorable, optimistic, or happy sentiment.}                                                      \\ 
\cline{2-4}
                    & \textit{Negative}     & \textcolor[rgb]{0.122,0.125,0.141}{[}0\textcolor[rgb]{0.122,0.125,0.141}{,}1\textcolor[rgb]{0.122,0.125,0.141}{]} & \makecell[lt]{The level of unfavorable, pessimistic, or unhappy \\ sentiment.}                                                 \\ 
\hline
\makecell{\textbf{Taboo} \\ \textbf{Rate}}
                    & \textit{Taboo}        & \textcolor[rgb]{0.122,0.125,0.141}{[}0\textcolor[rgb]{0.122,0.125,0.141}{,}1\textcolor[rgb]{0.122,0.125,0.141}{]} & \makecell[lt]{The level of potentially offensive or inappropriate \\ content expressed.}                          \\
\hline
\makecell{\textbf{TextBlob} \\ \textbf{Subjectivity}}       & \textit{Subjectivity} & \textcolor[rgb]{0.122,0.125,0.141}{[}0\textcolor[rgb]{0.122,0.125,0.141}{,}1\textcolor[rgb]{0.122,0.125,0.141}{]} & \makecell[lt]{The degree of personal opinion, bias, or perspective \\ expressed.}                                 \\ 
\hline
\end{tabular}
\end{table}

\subsection{Clustering users' psycholinguistic profiles}
In the following, we dive into the technical details of methods leveraged to obtain clusters starting from individual users' psycholinguistic profiles as well as to analyze results.
\\ \ \\
\noindent{\textbf{Feature Selection.}}
Given the extracted text features, we perform some basic statistical analysis to ensure their robustness with respect to the clustering task. At first, we study feature correlations in such a way as to avoid carrying redundant or similar information that could impact both the output and the execution time of the clustering algorithm. 
Relying on Pearson correlation coefficients, a widely used metric for quantifying linear relationships, we investigate the interplay of these features. Further, we assess the statistical significance of obtained correlations by computing p-values ($p<0.001$).  
Our analysis reveals that extracted features exhibit low statistically significant correlations, with an exception for the \textit{Valence} and \textit{Dominance} features that display the highest but still moderate positive correlation (i.e., 0.65) and for \textit{Subjectivity} and \textit{Sadnesss} that present a p-value higher than the imposed threshold. This outcome underlines their relevance and non-redundancy. Additionally, to assess the significance of extracted features in our research context, we also perform variance analysis. In detail, this step aims to identify and remove those features that lack meaningful variation, i.e., constant or quasi-constant variables. As for correlation analysis, all the features were preserved since they show a sufficient variance.
\\ \ \\
\noindent{\textbf{Clustering - K-Means.}} Clustering refers to the unsupervised task of grouping similar data points based on their features or characteristics. Among the existing clustering algorithms, we decided to rely on K-means \cite{macqueen1967classification} mainly for its interpretability and ability to deal with large datasets with a great number of data points (i.e., in our case 501,811).
K-means is a popular unsupervised algorithm for clustering/grouping unlabeled data points into distinct subgroups or clusters. The algorithm works by partitioning the data points into $K$  clusters, where each data point belongs to exactly one cluster. The number of clusters $K$ must be specified before running the algorithm. More in detail, K-means is iterative and starts by initializing $K$ cluster centroids at random or using a specific initialization method. Then, each data point is assigned to the nearest centroid based on a distance metric, and subsequently, centroids are updated by taking the mean of all the observations in each cluster. The algorithm repeats these steps until convergence, which occurs when the assignments no longer change or a maximum number of iterations is reached.
 \newline
In this work, we apply K-Means to the whole set of users' longitudinal psycholinguistic profiles (i.e., vectors of floats representing extracted textual features) in such a way as to investigate the transition of users in different clusters over time. To put it another way, a user who contributes in two different months is seen as two distinct data points and could be grouped into two different clusters based on the psycholinguistic profiles that describe him in a particular month. To choose the optimal number of clusters $K$, we rely on the Elbow method that consists of \textit{i)} varying the value of $K$ (from 2 to 10 in our case); \textit{ii)} computing for each value of $K$ the sum of the squared distance between each point and the centroid in a cluster; \textit{iii)} identify the ``elbow point``, i.e., the value of $K$ at which the sum of the squared distance starts decreasing at a slower rate. This point represents a trade-off between closely fitting the data and preventing unnecessary clusters. 
Our analysis reveals that the optimal cluster number for our work is 4.
\\ \ \\
\noindent{\textbf{Topic Modeling - BERTopic.}}
Topic modeling refers to algorithms tailored to uncover relevant patterns and topics within a collection of text documents. Accordingly, each document is represented as a mixture of topics and each topic as a distribution of words.
Among the various topic modeling implementations, we rely on BERTopic\footnote{\url{https://maartengr.github.io/BERTopic/index.html}} \cite{grootendorst2022bertopic} for its improved ability to capture more varied shades of meaning. 
BERTopic is a recently proposed topic modeling framework that - in its original implementation - uses the BERT model \cite{devlin2018bert} to transform words and documents into real-value vectors (i.e., word embeddings), reduces the number of features in the embeddings by leveraging UMAP \cite{mcinnes2018umap}, and clusters the documents into topics using as clustering algorithm HDBSCAN \cite{mcinnes2017hdbscan}. 
 \newline
In the current work, we apply BERTopic separately for each psychological cluster detected to retrieve and compare the most relevant themes that characterize online discussions. Regarding text preprocessing, in addition to adopting the same standard pipeline explained previously (see Section \ref{sez:feature_extr}), we lemmatize each word to obtain their root forms, which helps reduce variations and standardize the language representation.
As for the specific implementation of BERTopic, the algorithm provides a wide range of choices for its single components; therefore, it was customized to fit our textual data better. First, we employ BERTopic with a unigram to trigram range (1, 3) to ensure the extraction of both individual words and bigrams and trigrams. 
Then, after testing different dimensionally reduction and clustering methods, we opt for those presented in the original implementation, namely UMAP, and HDBSCAN. Specifically, we set a minimum cluster size of 200 to avoid small and noisy topics. Further, since using HDBSCAN might lead to the labeling of several documents as outliers (i.e., ~30\% in our work), we opt for a strategy to reduce/remove them. In detail, we compute the c-TF-IDF representations of outlier documents and assign them to the best matching c-TF-IDF representations of non-outlier topics. In the end, we obtain 10 topics for each cluster.
We show the distribution of retrieved topics within each depression cluster using bar charts  (see Fig. \ref{fig:all_clusters}(a-d top)). This visualization method allows us to identify the most dominant themes in each cluster and assess their prevalence. Also, we show the hierarchical representation of extracted topics in such a way as to study the relationships, overlaps, and similarities between different topics in each cluster (see  Fig. \ref{fig:all_clusters}(a-d bottom)). Finally, in the main article, Fig. 2(B) shows a scatter plot visualization that summarizes the profile clusters and the most frequent topics discussed in each cluster colored by the average NRCLex Valence score of the words included in that topic.



\section{Networks and Transition Patterns}
\label{sec:interacts}

We estimate exposures to clusters by quantifying interactions between users.
Exposures assess the level of contacts/interactions occurred between class $X$ and class $Y$.
Leveraging different graph- and hypergraph-based notions of users' neighbourhood, we can have access to different strengths of exposure.
\\ \ \\
\noindent \textbf{Graph notions.}
Formally, a graph $\mathcal{G}=(V,E)$ is composed of a set $V$ of nodes and a set $E$ of edges which are unordered pairs of elements of $V$ \cite{newman2018networks}.
The neighbourhood of a node $u$, denoted as $\Gamma(u)$, is the set of $u$'s adjacent nodes, namely the nodes with whom $u$ has an explicit edge.
Clusters can be represented as an attribute $a=\{l_1,...,l_d\}$, where $l_i$ is a cluster label id, such that $a(u)$, with $u \in V$, identifies the cluster label id associated to $u$.
Hence, the node-attributed graph neighbourhood of a node $u$ is a list of cluster labels associated with the nodes in $\Gamma(u)$, i.e., $\Gamma_a(u) = [a(v)|v \in \Gamma(u)]$.
To strengthen the exposure to a cluster label (see ``Majority of Interactions" exposure, next), we can compute the most frequent label within $\Gamma_a(u)$.
The relative frequency of a cluster label $l_i$, i.e., $f(l_i)$, is the number of times $l_i$ occurs in a list, divided by the total frequency.
The most frequent cluster label within $u$'s node-attributed neighborhood is denoted as $argmax_{l_i \in \Gamma_a(u)}(f(l_i))$.
\\ \ \\
\noindent \textbf{Hypergraph notions.}
A hypergraph $\mathcal{H}=(V_H, E_H)$ is composed of a set $V_H$ of nodes and a set $E_H$ of non-empty subsets of $V$ called hyperedges \cite{battiston2020networks}.
In a hypergraph, the neighbourhood of a node $u$ is the set of hyperedges that contain $u$. We denote it as $\Gamma_H(u)$.
It is interpreted here as the set of subthreads/discussions where the target user was involved during its permanence on the Reddit platform.
Again, we leverage node-attributed information as follows: Clusters are an attribute $a=\{l_1,...,l_d\}$, with $l_i$ labels, such that $a(u)$, with $u \in V_H$, denotes $u$'s cluster label id.
However, differently from graphs, node-attributed hypergraph neighbourhoods are built upon ``characteristic" values of node-attributed hyperedges.
A node-attributed hyperedge is a list of cluster labels associated to the members of hyperedge, i.e., $e_{H_a} = [a(v)|v \in e_H]$, with $e_H \in E_H$.
We can sum up node-attributed hyperedges with a single, ``characteristic" value, e.g., the most frequent label within the list, namely $argmax_{l_i \in e_{H_a}}(f(l_i))$, where $f(l_i)$ is the relative frequency of a cluster label $l_i$ in the node-attributed hyperedge $e_{H_a}$.
Hence, we can build the list that contains only the most frequent labels within the hyperedges that contain a target node $u$: This is the node-attributed hypergraph neighbourhood of the node $u$. We denote it as $\Gamma_{H_a}(u)$.
When computing the presence of ``at least one" cluster label in the node-attributed hypergraph neighbourhood of a target node, the interpretation is that we count a cluster label which is maximal within the subthread/discussion where the target node was involved.
Also, strengthening the exposure to a target cluster label means to compute the most frequent ``characteristic" label within $\Gamma_{H_a}(u)$, namely $argmax_{l_i \in \Gamma_{H_a}(u)}(f(l_i))$, where $f(l_i)$ is the relative frequency of a ``characteristic" cluster label.
\\ \ \\
\noindent \textbf{Social Exposure}.
We aim to quantify the exposure of a node $u$ to a target cluster label $l_i$ in $u$'s node-attributed neighbourhood.
From the previous formalization, it is still missing the role of the temporal dimension, which is actually the fundamental element to understand the \textit{consecutio temporum} of the ``exposures" because users' labels and interactions are captured month by month.
Hence, for each month, we ``capture" a snapshot of interactions and a cluster label for each user.
We can leverage the snapshot-based representation for temporal networks as follows: A snapshot graph $\mathcal{G}_T$ is an ordered set $\langle \mathcal{G}_1, ..., \mathcal{G}_t \rangle$ where each snapshot $\mathcal{G}_i=(V_i,E_i)$ is univocally identified by the sets of nodes $V_i$ and edges $E_i$ \cite{rossetti2018community}.
The definition is identical for hypergraphs.
 \newline
As described in the main article, we test four different levels of exposure as in the following:

\begin{enumerate}[i]
    \item {\bf Single Interaction}: There exists, at $\mathcal{G}_{t-1}$, at least one $u$'s neighbor having as a cluster label $l_i$, i.e., $\exists\ a(v)=l_i$, with $a(v) \in \Gamma_a(u)$;
    \item {\bf Single Homogeneous Context}: There exists, at $\mathcal{H}_{t-1}$, at least one $u$'s neighbor having as a ``characteristic" cluster label $l_i$, i.e., $\exists\ a(v)=l_i$, with $a(v) \in \Gamma_{H_a}(u)$;
     \item {\bf Majority of Interactions}: At $\mathcal{G}_{t-1}$, $l_i$ is the most frequent cluster label within $u$'s neighbors, i.e., $argmax_{l_i \in \Gamma_a(u)}(f(l_i))$;
   \item {\bf Majority of Homogeneous Contexts}: At $\mathcal{H}_{t-1}$, $l_i$ is the most frequent cluster label within $u$'s neighbors, i.e., $argmax_{l_i \in \Gamma_{H_a}(u)}(f(l_i))$;.
\end{enumerate}

Transitions are counted only if the constraint on the exposure is true.
\\ \ \\
\noindent \textbf{Null Models}.
Transitions require statistical validation to take into account the effects of the distribution of cluster profiles and the temporal dependency on the transitions.
Fig. \ref{fig:pers_and_distr}(a) shows that the distribution of cluster labels is imbalanced during the whole observation period.
The so-called in-between profile ($C4$) is the most frequent one, while the mildly positive one ($C3$) is the minoritarian one.
Hence, it is important to test the impact of such imbalanced distributions on the transitions.
We build a null model where transitions are computed on a replica of the original dataset where users' cluster labels are randomly shuffled.
To build the model, the shuffling operation is repeated many times.
We refer to this model as the \textit{Cluster Null Model}.
It outputs a mean and a standard deviation for each probability of transition between cluster X to cluster Y in order to compare these expected values to the original probabilities of transition.
 \newline
Also, the temporal dimension -- the time when user interactions occur --- is an important factor to consider to test the validity of transitions.
Several randomizing strategies for temporal networks are available in the literature \cite{gauvin2018randomized}.
For us, it is important testing the impact that the order of interactions has had on the transitions.
Following this rationale, we build a null model where transitions are computed on a replica of the dataset where graph/hypergraph snapshots are randomly shuffled.
We refer to this model as the \textit{Temporal Null Model}.
Also, this model outputs a mean and a standard deviation to compare with the original values.
 \newline
To evaluate whether the transitions are statistically significant, we compute the following z-score:

\begin{center}
    $z= \frac{x - \mu}{\frac{\sigma}{\sqrt n}},$
\end{center}

where $x$ is the probability of shifting between cluster X and cluster Y in the original dataset, $\mu$ is the expected probability when cluster labels (Cluster Null Model) or snapshots (Temporal Null Model) are randomly shuffled, $\sigma$ is the standard deviation of such ensemble, and $n$ is the number of times the process is repeated.
\\ \ \\
\noindent \textbf{Users' persistence}.
The main focus of this work was analyzing transitions between clusters to study the effects of online social interactions.
As specified in the main article, to analyze the transitions from one cluster to another, we retained only the users who had participated in discussions for at least two consecutive months.
Fig. \ref{fig:pers_and_distr}(b) shows the distribution of users' permanence during the two-year-long observation period.
In the plot, it is not reported the volume of ``discarded" users, namely those who appeared in only one month, which is approximately $150.000$ users corresponding to $\simeq 50\%$ of the entire population.
The plot shows that $>40.000$ users (among $\simeq 150.000$ filtered users) participated in depression-related discussions for exactly two months, then the long-tail distribution highlights a decreasing number of users, such that only a few users actively participated during the whole span of the observation period.
\\ \ \\
\noindent \textbf{Notes on the interpretation of Transition Patterns}.
Probabilities are normalized with respect to the rows (matrices) and the out-going flow (Markov chains), such that the sum of the values on the rows/out-going edges from a cluster to all other clusters is 1.
These probabilities map the emotional footprint of groups of users active in mental health discussions through their ``orientations" over time.
Hence, self-loops can be interpreted as a tendency of a specific ``emotionally-profiled" group of users to not swing between different emotional states in a specific mental health-related discussion on the platform.
High probabilities of remaining in the same cluster may indicate groups of users that tend to be emotionally stable with respect to a theme.
Using different users' interaction levels of exposure helps us to quantify the effect that the exposure to a specific cluster has on the transitions observed in the future.
\\ \ \\
\noindent \textbf{Transition Patterns on the Metapopulation}.
In the main article, we focused on Constrained Transition Patterns (CTP), where the ``constraints" are related to the conditional probability of observing individuals' transition from one cluster to another, given interaction exposures to a target cluster.
  
However, Transition Patterns (TP) can also be observed by ``switching off" the constraints on users' interactions, thus observing users' evolution over time without imposing to characterize the role that interactions may have on eventual shifts.
Fig. \ref{fig:metapop} shows statistically significant TP between emotional states on the metapopulation.
The statistical test has been done destroying users-profiles associations only, not having temporal dependency of interactions to observe.
Statistically significant transitions occur for the positive profile ($C2$) and the mildly positive profile ($C4$) only.
In particular, members of the positive profile ($C2$) are likely to mildly ($C3$, $13\%$) or strongly ($C1$, $17\%$) worse their conditions or to prefer a shift towards the in-between cluster ($C4$, $39\%$).
Members of the mildly positive profile ($C4$) can worse their conditions ($C1$, $18\%$), or be found in the in-between cluster ($C4$, $42\%$). 
\\ \ \\
\noindent{\textbf{Constrained Transition Patterns}.}
In the main article, we have shown results about Constrained Transition Patterns (CTP) by keeping transitions that are statistically significant with respect to both null models.
Fig. \ref{fig:dep_constr_1} and Fig. \ref{fig:dep_constr_2} show results by keeping significant transitions with respect to the single null models.
Generally, by destroying the temporal dependence of interactions, we observe a larger number of statistically significant transitions than destroying the cluster volumes only, meaning that the majority of transitions that are not considered strongly depend on the volume of clusters.
 \newline
From the first row of Fig. \ref{fig:dep_constr_1}, we can observe that conditioning over $C1$ continues to generate statistically significant transition only for users in $C3$.
However, here we observe an aggravation of their conditions also in case of single interactions (to $C1$), a link which is not kept by considering both models.
Also, we observe an improvement of their conditions in the case of single homogeneous contexts (to $C2$), which is not kept by considering both models.
From the first row of Fig. \ref{fig:dep_constr_2}, more transitions are kept, also for the majority of interactions/homogeneous contexts. 
It is interesting to observe that -- when only the temporal dependency is destroyed -- conditioning over $C1$ generates statistically significant transitions also for the users in $C2$ in case of single and majority of interactions, and statistically significant transitions for all the other three profiles in case of single and majority of homogeneous contexts, particularly towards $C1$ itself.
This means that being exposed to $C1$-related contexts rather than pairwise interactions drastically worsens the condition of all the other groups.
 \newline
The second row of Fig. \ref{fig:dep_constr_1} shows quite similar results to those presented in the main article, indicating that $C3$ and $C4$ profiles are the most affected by interactions with the positive members of $C2$.
Interestingly, from the second row of Fig. \ref{fig:dep_constr_2}, we observe that being exposed to $C2$-related single interactions and single homogeneous contexts can improve the conditions of all three other profiles.
However, there are no links from $C1$ to $C2$ in the case of neither the majority of single interactions nor the majority of homogeneous contexts, perhaps indicating that ``strong" exposures to positive users may not lead to desired improvements.
 \newline
When considering both null models, from the third row of Fig. \ref{fig:dep_constr_1}, we observe that the statistically significant transitions conditioned by interactions with users in $C3$ depend on the Cluster Null Model.
When destroying the temporal dependency only (Fig. \ref{fig:dep_constr_2}), we obtain more transitions, however hard to discuss.
$C3$ is a minoritarian cluster of mildly positive users.
Conditioning over them generates transitions from $C2$ to $C3$ except in the case of the majority of single interactions and transitions from $C1$ to $C3$ except in the case of single homogeneous contexts.
However, being $C3$ able to affect a relatively small fraction of the overall population, e.g., $4\%$ in case of the majority of interactions, or $3\%$ in case of the majority of homogeneous contexts, we do not find such transitions relevant to describe and characterize a population of users discussing depression-related topics.
 \newline
Also, the fourth row of Fig. \ref{fig:dep_constr_1} describes the same statistically significant transitions as the ones described using both null models.
Detailed interpretations about them are left to the main article.
From the fourth row of Fig. \ref{fig:dep_constr_2}, we also observe that the positive members in $C2$ are likely to transit to $C4$ when exposed to them, except in the case of the majority of single interactions.
In the case of the majority of homogeneous contexts, all the other three profiles tend to transit to $C4$, having participated in interactions/discussions with members in $C4$.

\begin{figure*}[h!]
    \centering
    \begin{subfigure}[b]{0.9\textwidth}
        \centering
        \includegraphics[width=\textwidth]{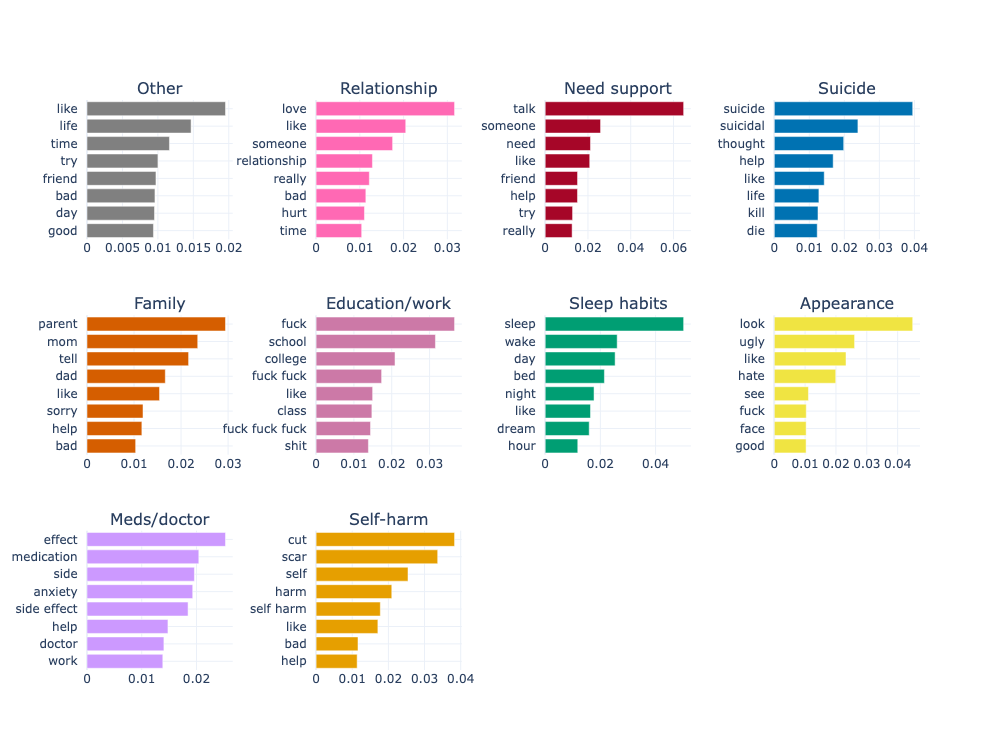}
        \caption*{a) Cluster 1}
        \label{fig:dep_cluster1}
    \end{subfigure}
\end{figure*}
\begin{figure*}[h!]
    \centering
    \begin{subfigure}[b]{0.9\textwidth}
        \centering
        \includegraphics[width=\textwidth]{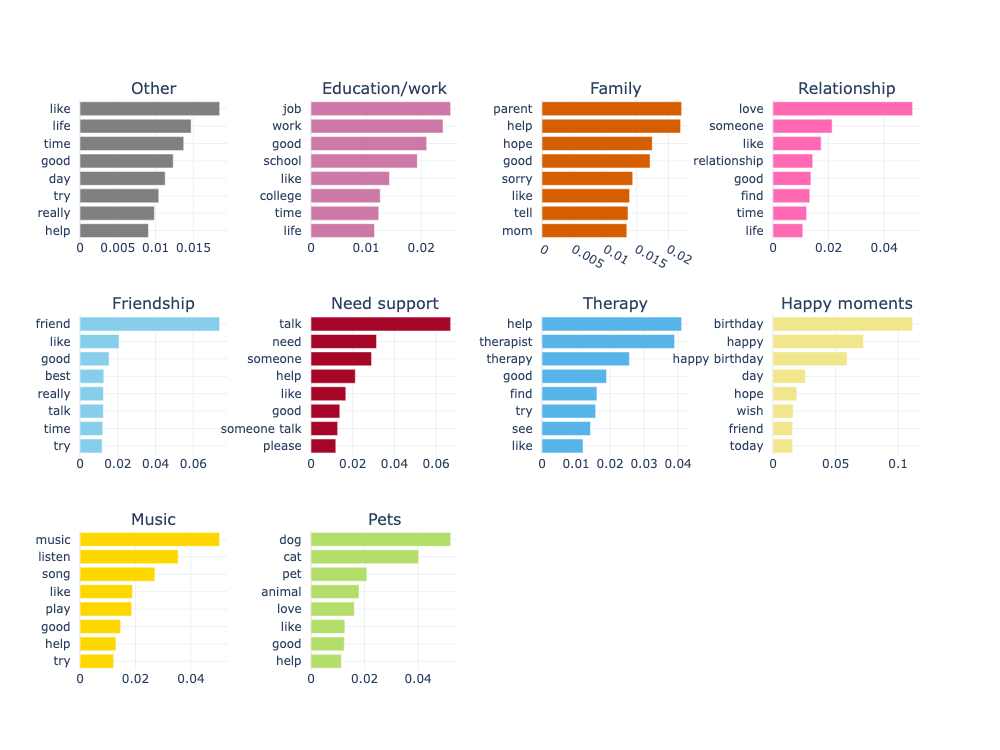}
        \caption*{b) Cluster 2}
        \label{fig:dep_cluster2}
    \end{subfigure}
\end{figure*}
\begin{figure*}[h!]
    \centering
    \begin{subfigure}[b]{0.9\textwidth}
        \centering
        \includegraphics[width=\textwidth]{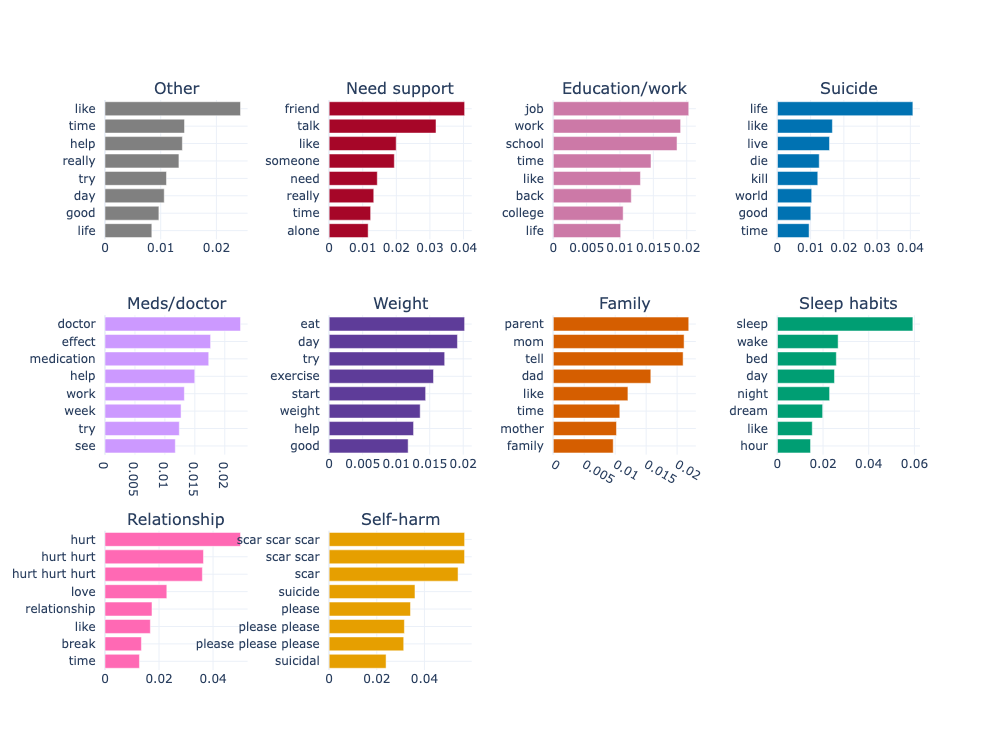}
        \caption*{c) Cluster 3}
        \label{fig:dep_cluster3}
    \end{subfigure}
\end{figure*}
\setcounter{figure}{0} 
\begin{figure*}[h!]
    \centering
    \begin{subfigure}[b]{0.9\textwidth}
        \centering
        \includegraphics[width=\textwidth]{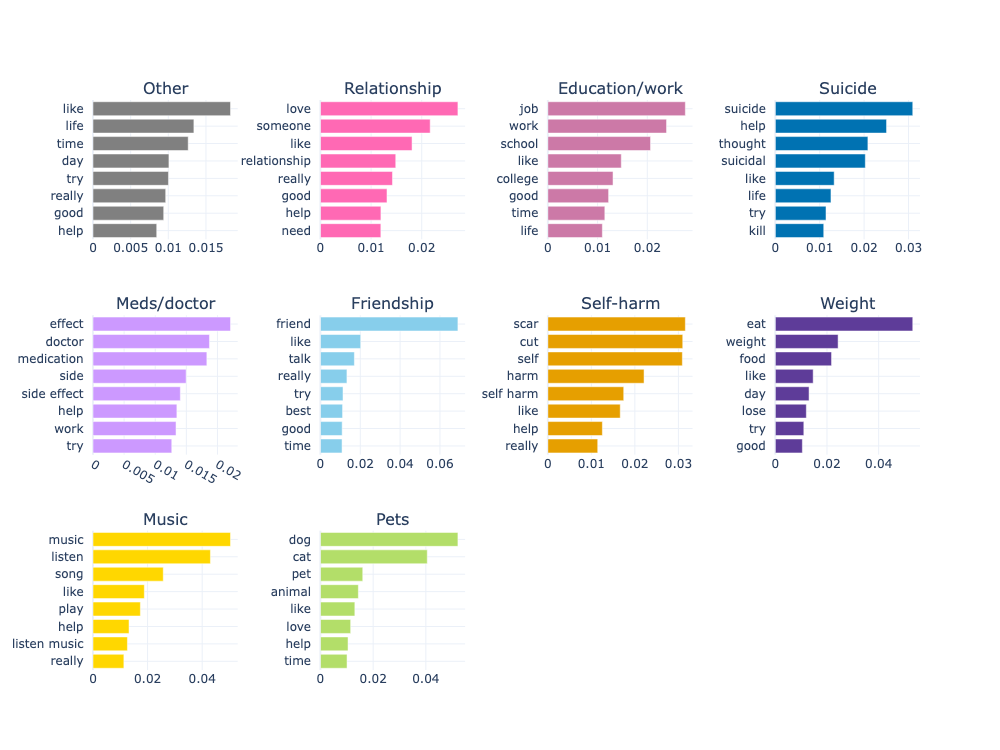}
        \caption*{d) Cluster 4}
        \label{fig:dep_cluster4}
    \end{subfigure}
     \caption{For each cluster (a-d), bar charts display the 10 most frequent topics (in descending order) extracted from users' texts using BERTopic. Each bar chart visualizes the frequencies of the top 8 representative words for each topic (based on TF-IDF) along the x-axis. These words can be unigrams, bigrams, or trigrams. Topics are labeled using single terms that describe their content, such as ``job", ``work", ``school", ``college" for \textit{Education/work} topic. Notice that we label as ``Other" those extracted topics (in grey) that are composed of words too general to be categorized with a label.}
    \label{fig:all_clusters}
\end{figure*}

\begin{figure}
    \centering
    \subfloat[]{\includegraphics[scale=0.4]{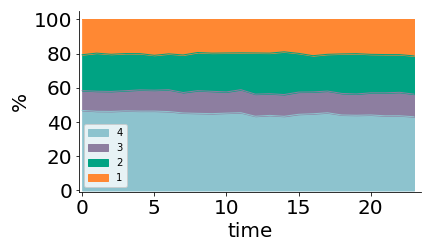}}
    \subfloat[]{\includegraphics[scale=0.4]{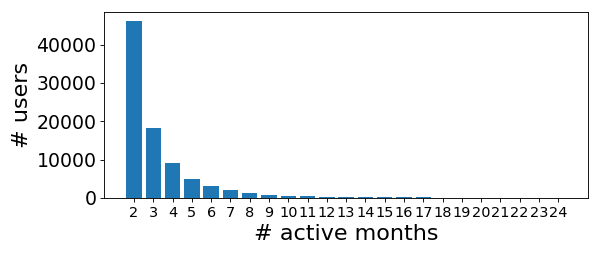}}
  \caption{(a) Distribution of the clusters over time; (b) Distribution of the user permanence: The number of active months on the x-axis indicates the number of snapshots in which users are active on the platform; }
  \label{fig:pers_and_distr}
  \end{figure}

\begin{figure}
    \centering
    {\includegraphics[scale=0.15]{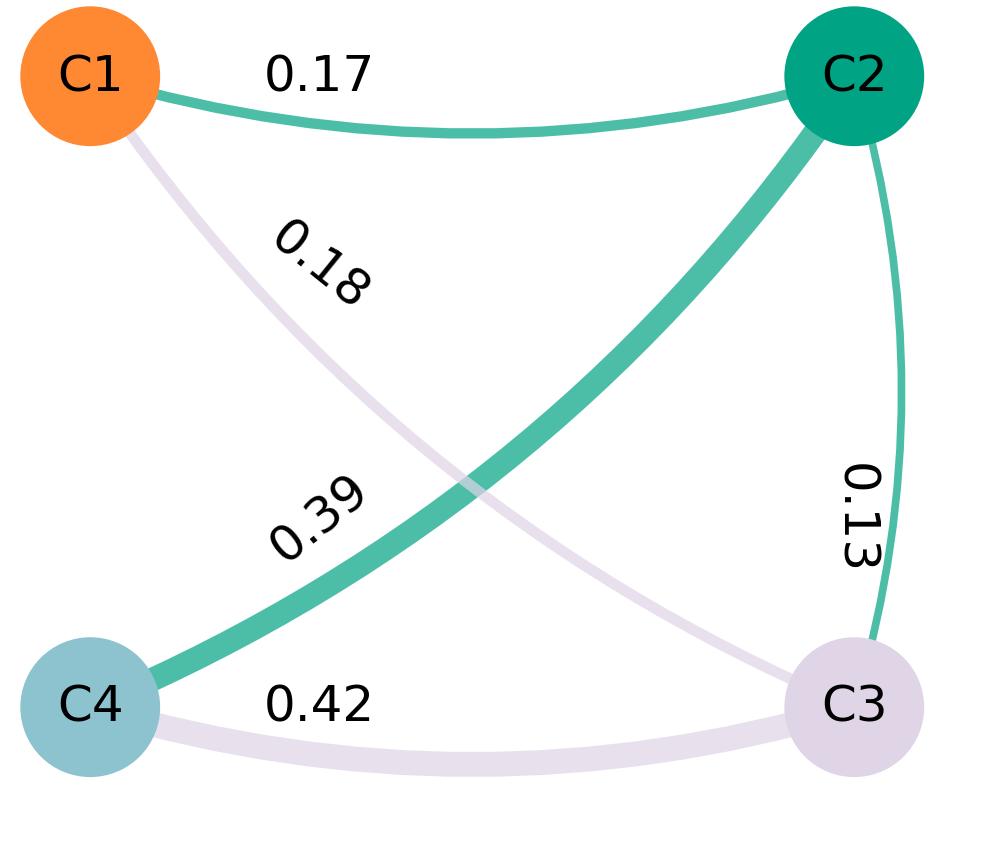}}
  \caption{Visual representation of Markov transition matrices among user clusters given the metapopulation only filtered based on volume statistical significance.}
  \label{fig:metapop}
  \end{figure}

\begin{figure}[h]
    \centering
    \includegraphics[scale=0.61]{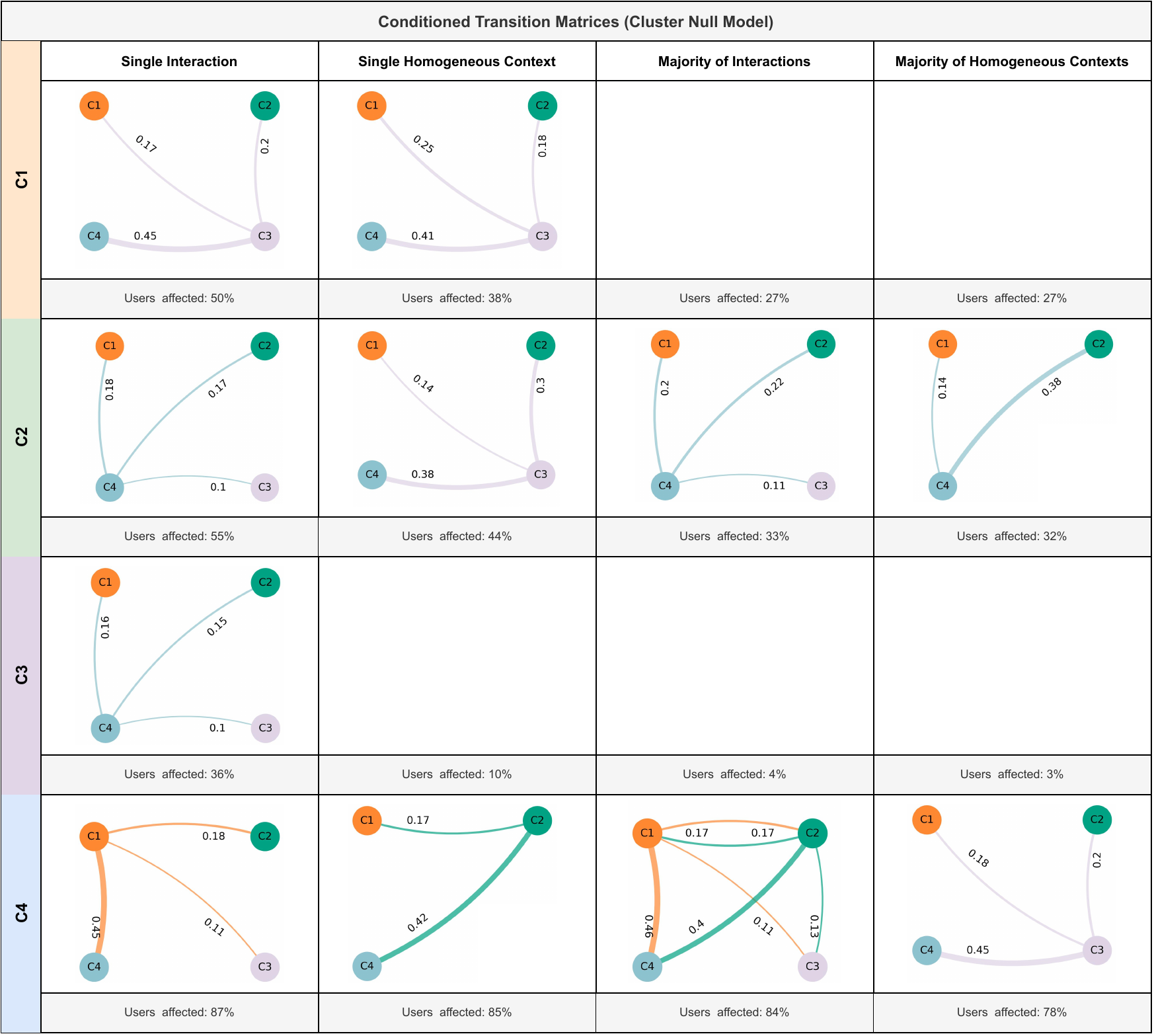}
    
  \caption{ Constrained Transition Patterns: Cluster Null Model}
  \label{fig:dep_constr_1}
  \end{figure}
  \begin{figure}[h]
    \centering
    \includegraphics[scale=0.61]{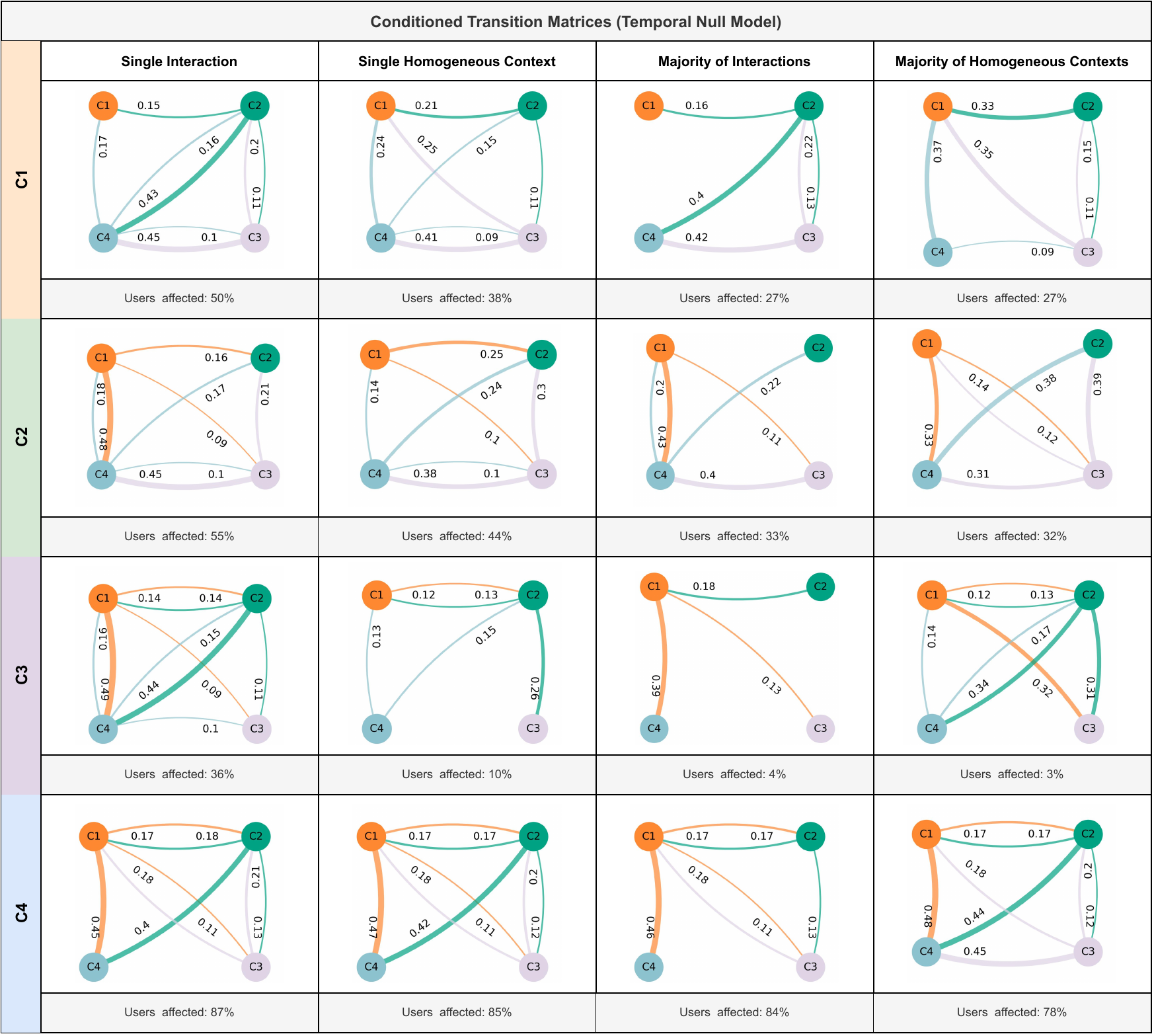}
    
  \caption{Constrained Transition Patterns: Temporal Null Model}
  \label{fig:dep_constr_2}
\end{figure}


\end{document}